\documentclass[showpacs,preprintnumbers,amsmath,amssymb,24pt]{revtex4}
\usepackage{graphicx}
\usepackage{graphics,epsfig}
\usepackage{bm}
\usepackage{amssymb}

\usepackage{amsmath}
\usepackage{subfigure}
\usepackage[bf]{caption}

\begin{document}

\title{Combining tomographic imaging and DEM simulations to investigate the structure of experimental sphere packings}
\author{
Gary W Delaney$^{1,2}$
T. Di Matteo$^{2,3}$,
and
Tomaso Aste$^{2,3,4}$,
}
\address{
$^1$ CSIRO Mathematical and Information Sciences, Private Bag 33, Clayton South, Vic, 3168, Australia
$^2$ Department of Applied Mathematics, School of Physical Sciences, The Australian National University, Canberra, ACT 0200, AUSTRALIA;\\
$^3$ Department of Mathematics,  King's College, The Strand, London, WC2R 2LS, UK;\\
$^4$ School of Physical Sciences, University of Kent, Canterbury, Kent, CT2 7NH, UK.
}


\begin{abstract} \baselineskip14.pt 
We combine advanced image reconstruction techniques from computed X-ray micro tomography (XCT) with state-of-the-art discrete element method simulations (DEM) to study granular materials. 
This ``virtual-laboratory'' platform allows us to access quantities, such as frictional forces, which would be otherwise experimentally immeasurable.
\end{abstract}

\maketitle 

\section{Introduction}
Everywhere we look in nature and the modern world we find granular matter. 
From sands to soils, grains to powders, our understanding of granular systems has applications in diverse areas of both fundamental and applied science \cite{Kadanoff1999}. 
In recent times, two key technologies that have been employed in attempting to gain a deep understanding of the properties of granular materials, namely, computational modelling using the Discrete Element Method (DEM) \cite{Cundall1979,Hutzler2004,ppp,Kadanoff1999,Jaeger96} and direct 3D imaging of experimental systems using X-Ray tomography (XCT) \cite{AstePRE05,Tsukahara2008}. 
From XCT we can obtain impressive experimental measurements of real physical systems, allowing us to measure the internal structural properties of a granular packing. 
However, some information such as frictional forces cannot be retrieved from a pure geometrical characterization of the system structure.
In this respect, DEM offers the ability to fully characterise the system under consideration, allowing access to all its properties down to the forces on individual grains and at inter-grain contacts. 
Here, we combine these two technologies to obtain the best advantages of both worlds, considering real experimental granular systems from which we can also obtain all the additional information available to us from DEM. 

In our approach, we use a tomographic reconstruction of a real sphere packing as our starting point, and input this experimental data (in the form of the coordinates of the sphere centers) into our DEM simulation \cite{Delaney2007,Delaney2008-1}. By this technique we can better characterize our experimental packing by removing the uncertainty in the exact locations of the sphere centers ($<$ $0.1\%$) and also by removing the uncertainty in the exact diameters of the spheres caused by the polydispersity ($\sim$~2\%). We thus produce a simulated ideal mono-disperse sphere packing that almost exactly matches the grain positions of the original experimental packing. From such a `virtual packing' we can now compute with precision several static and dynamical properties which are otherwise not directly accessible from experiments.

\section{Experiments}

We use the AAS database on disordered packings \cite{Database} which contains structural data from experimental sphere packings obtained by X-ray Computed Tomography. 

We analyze 6 samples (A-F) composed of acrylic beads prepared in air  within a cylindrical container with an inner diameter of $ 55\; mm$ and filled  to a height of $\sim 75\; mm$ \cite{AstePRE05,AsteKioloa,Aste05rev}.  
Samples A and C contain $\sim 150,000$ beads with diameters $d =1.00\;  mm $ and polydispersity within $0.05 \; mm$.
Whereas, samples B, D-F contain $\sim 35,000$ beads with diameters $d = 1.59 \; mm $ and polydispersity within $0.05 \; mm$.
The two samples at lower packing fraction (A, B) were obtained by placing a stick in the middle of the container before pouring the beads and then slowly removing the stick \cite{ppp}. 
Sample C  was prepared by gently and slowly pouring the spheres into the container.
Whereas, sample D was obtained by a faster pouring. 
In sample E, a higher packing fraction was achieved by gently tapping the container walls.  
The densest sample (F)  was obtained by a combined action of gentle tapping and compression from above (with the upper surface left unconfined at the end of the preparation). 
To reduce boundary effects, the inside of the cylinder was roughened by randomly gluing spheres to the internal surface.
The geometrical investigation of the packing structure was performed over a central region at 4 sphere-diameters away from the sample boundaries. 
The packing fraction of each of the samples is: A, $\rho \sim 0.586$; B, $\rho \sim 0.596$; C, $\rho \sim 0.619$; D, $\rho \sim 0.626$; E, $\rho \sim 0.630$; and F, $\rho \sim 0.640$. 

We also analyze 12 samples (FB12-24 and FB27) containing about 150,000 glass beads with diameters $0.25\;mm$ placed in a vertical polycarbonate 
tube with an inner diameter of 12.8 $mm$ and a length of 230 $mm$.
The packings were prepared in water by means of fluidized bed technique \cite{Database,Schroder05,AsteEPL07}. 
Packing fractions between 0.56 and 0.60 were obtained by using different flow rates. 
After each flow pulse, the particles sedimented forming a mechanically stable packing. 
Packings created in this way are in a stationary state with packing fractions which are independent of the preparation history and have average values which depend on the flow rate, with smaller packing fractions for higher flow rates (larger bed expansions).

We use X-ray Computed Tomography (XCT) to calculate the beads centers coordinates.
This is done by applying  over the three-dimensional XCT density map a convolution method which is made algorithmically very efficient by using (parallel) Fast Fourier Transform.
Furthermore a watershed method is also used to identify distinct grains.

The resulting beads positions are estimated with a precision better than 0.1\% of the beads diameters.
This precision is well below the beads polydispersity which is between 1\% and 3\% considering both changes in the bead sizes and deviations from perfect sphericity.

\section{Simulation}

\begin{figure}
\centering
\includegraphics[width=0.49\columnwidth]{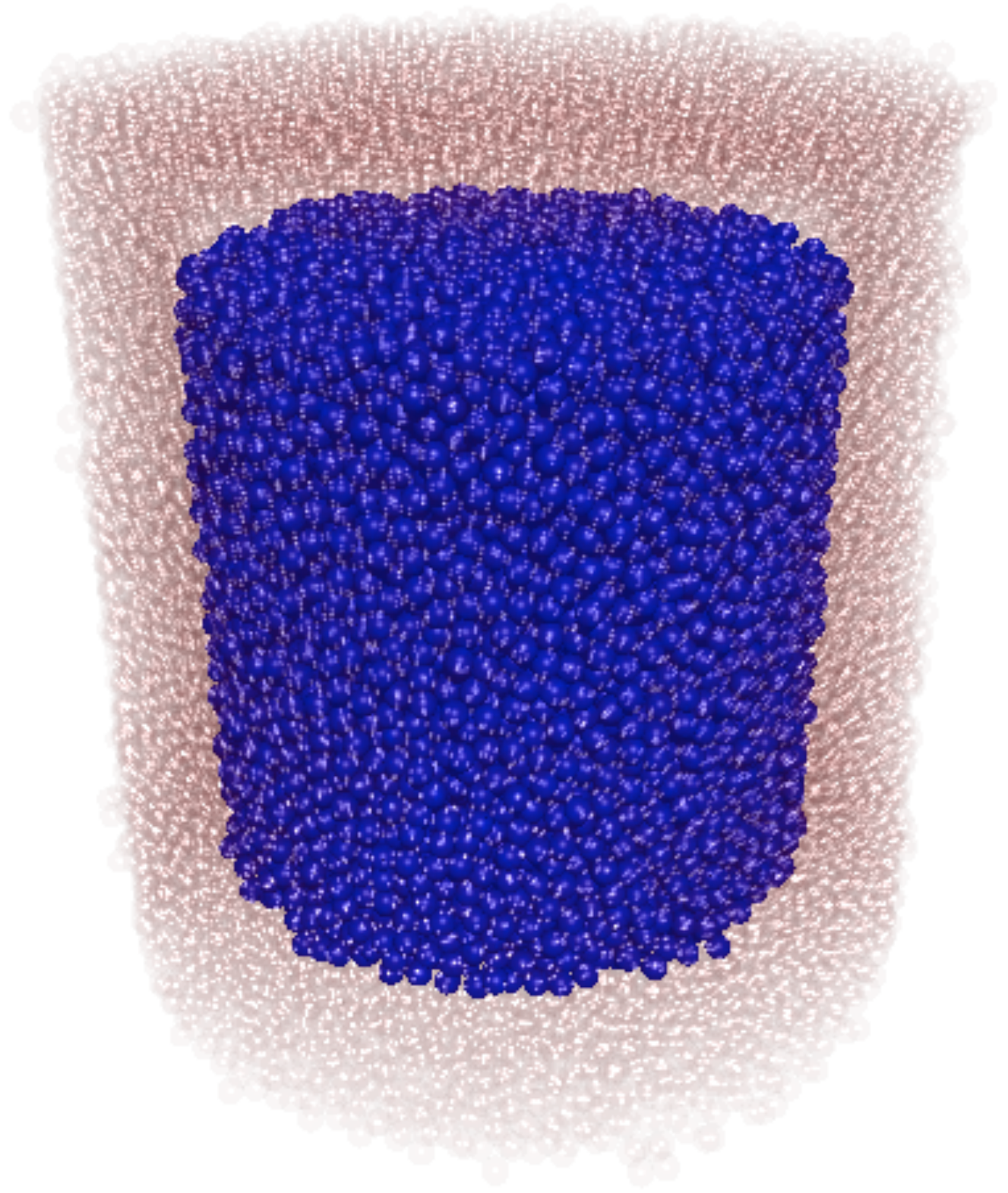}
\includegraphics[width=0.49\columnwidth]{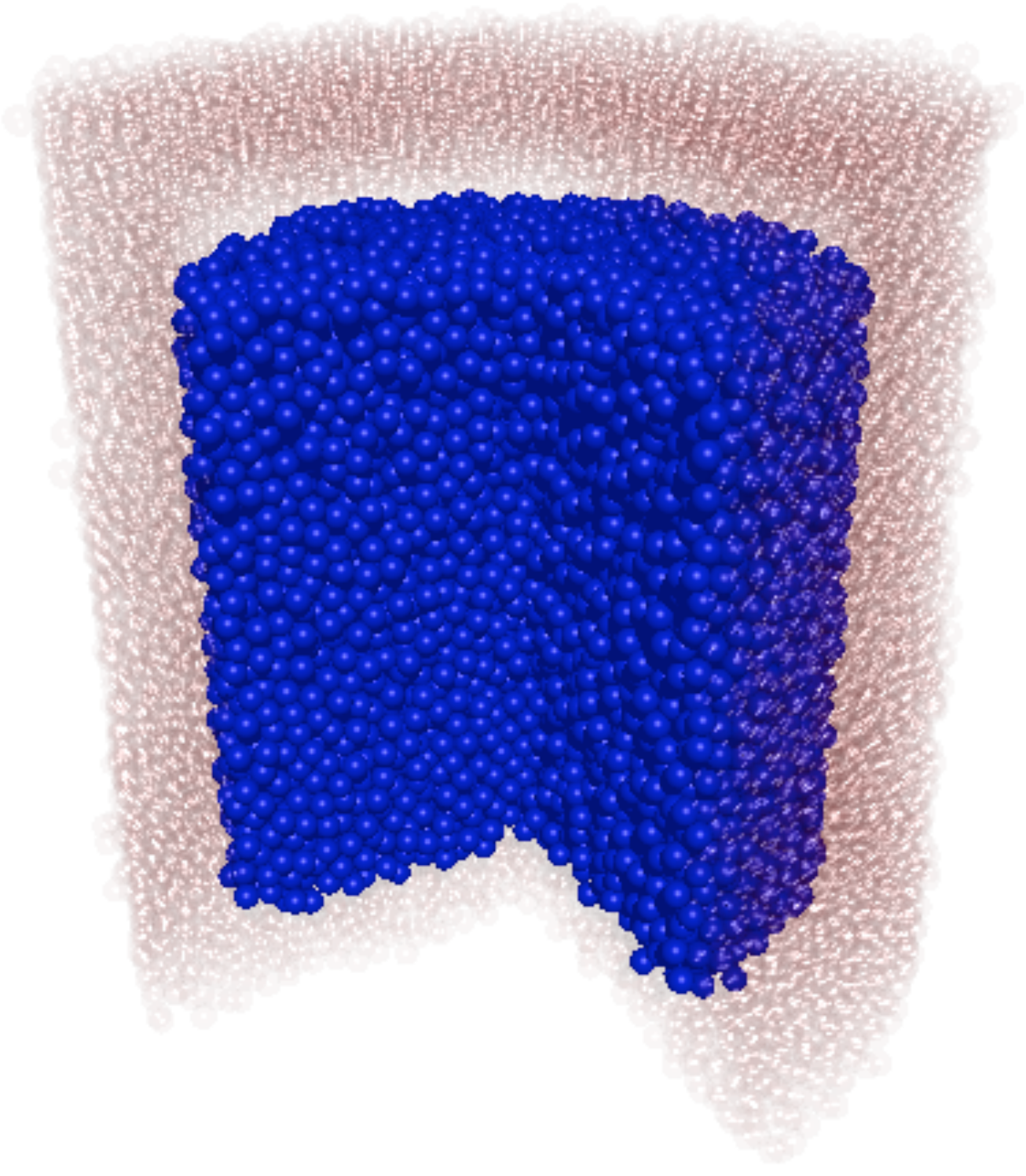}
\caption{ 
Combined tomographic and DEM reconstruction of a sphere packing. The inner spheres that are part of the DEM simulation are shown in blue, with the fixed boundary spheres rendered transparent. Images are for the F sample with 6856 spheres in the internal region. The entire sample is shown (left) and also the sample with one quarter removed (right) to show the internal structure of the packing. 
}
\label{f:height}
\end{figure}

In order to eliminate the unavoidable experimental incertitudes on the bead positions, sizes and shapes we reconstruct a numerical sample by using the experimental coordinate and gently relaxing the system to perfect spherical beads. 
The resulting `virtual packing' has a geometrical structure that is almost identical to the experimental one. 
To this purpose we use DEM simulations that integrate Newton equation of motion for viscoelastic grains with rotational and translational degrees of freedom. 
In a typical process, grains interact during collision with non-linear forces and energy is continuously dissipated by inelastic collisions, by friction and by viscous damping.

Our simulations consider mono-disperse packings of elastofrictional spheres
that interact with a normal repulsive force $F_n$ proportional to their overlap $\xi_n$ accordingly with the Hertz's formula 
\begin{equation}
 F_n = k_n \xi_n^{3/2}
\end{equation}
with $\xi_n = d - |{\vec{r}_i} - {\vec{r}_j}|$, where $d$ is the sphere-diameter and  $\vec{r}_i$ and $\vec{r}_j$ are the positions of the grain centers \cite{Landau1970,Makse04,Hutzler2004}. This force only acts when grains overlap, with $F_n = 0$ for $\xi_n < 0$. 
The spring constant $k_n$ is related to the physical properties of the grains, with:
\begin{equation}
 k_n = \frac{\sigma^{1/2}E}{3(1-\nu^{2})},
\end{equation}
where $E$ is Young's modulus and $\nu$ is Poisson's ratio. 
We also consider tangential force $F_t$ under oblique loading using  
\begin{equation}
 F_t = -\textrm{min}(|k_t \xi_n^{1/2} \xi_t |, |\mu F_n |) \cdot \textrm{sign}(v_t)
\end{equation}
where $\xi_t$ is the displacement in the tangential direction that has taken place since the time $t_0$ when the two spheres first got in contact 
\begin{equation}
 \xi_t =  \int_{t_0}^t v_t (t') \, dt'. 
\end{equation}
with $v_t$ is the relative shear velocity and $\mu$ is the kinematic friction coefficient between the spheres \cite{Cundall1979}.  
Furthermore, we include a normal viscoelastic dissipation  
\begin{equation}
 F_n = - \gamma_n \xi_n^{1/2} \dot{\xi}_n
\end{equation}
where $\dot{\xi}_n$ is the normal velocity, and a viscous friction force
\begin{equation}
F_t = - \gamma_t v_t
\end{equation}
\cite{Schafer96}. 

\section{DEM Relaxation of experimental samples}

The DEM simulation is performed on a set of unfixed spheres in the central region of the sample, with the boundaries provided by the outer spheres  which are kept fixed. 
The simulation uses realistic physical parameters with acrylic beads having: Young modulus $3.2 \; GPa$; Poisson ratio 0.3; density of $1150 \;kg/m^3$; inter grain static friction coefficient 0.28.
Samples A,C have radius 0.5 mm whereas  samples B,D,E,F have radius 0.795 mm.
The glass beads have: Young modulus $70 \; GPa$, a Poisson ratio 0.2; density of $2500 \;kg/m^3$; inter grain static friction coefficient 0.9 and radius 0.125 mm. 
Figure \ref{f:height} shows the variation in the average height of the grains as the simulation proceeds. 
Typically during the DEM relaxation there is initially a small expansion of the system, with the height increasing by a fraction of percent of the initial eight. 
The viscoelastic dissipation and viscous friction forces remove energy from the system until the spheres have all settled down to a stationary state. 
The final average height of the grains is within 0.1\% - 0.2\% of the initial average height and the average displacement of the centers of the spheres during the relaxation process is less than 5\% of the sphere diameters. 

The relaxation is performed using large values of the the normal and tangential dissipation parameters $\gamma_n$ and $\gamma_t$. This ensures a purely quasi-static relaxation, with the grains having during all the relaxation process an average  number of contacts larger than 4.
Essentially, the relaxation dynamics consists in a re-equilibration of the elastic stress trapped in the grain overlaps in the initial configurations which are unphysical and unbalanced due to the experimental uncertainties on the grain positions, polydispersity and a-sphericity.

 \begin{table}
 \caption{\label{t.1}
Details for the fluidized bed in the internal region of the DEM relaxed samples:
Number of internal mobile spheres (N), Packing fraction ($\phi$), Average number of neighbors ($z_c$), Fraction of contacts that cannot hold extra tangential force ($q$), Prediction for the isostatic condition from Eq.\ref{ISO} ($z_{ISO}$), critical size ($l^*$) in units of $d$.
}
\begin{ruledtabular}
 \begin{tabular}{lllllll}
 Sample & $N$ & $\phi$      &  $z_c$       &   $q$    &  $z_{ISO}$ &  $l^*$   \\ \hline
FB18 & 	6857 & 	0.5738 &	4.70 & 	0.026 & 	4.04 & 	11.8  \\ 
FB17 & 	7242 & 	0.5756 &	4.69 & 	0.027 & 	4.04 & 	10.7  \\ 
FB15 & 	7041 & 	0.5778 &	4.71 & 	0.025 & 	4.03 & 	10.3  \\ 
FB14 & 	6922 & 	0.5780 &	4.70 & 	0.027 & 	4.04 & 	9.2  \\ 
FB20 & 	6746 & 	0.5785 &	4.69 & 	0.022 & 	4.03 & 	9.4  \\ 
FB16 & 	6804 & 	0.5806 &	4.71 & 	0.029 & 	4.04 & 	10.4  \\ 
FB19 & 	6856 & 	0.5824 &	4.72 & 	0.030 & 	4.04 & 	11.6  \\ 
FB21 & 	6927 & 	0.5851 &	4.74 & 	0.027 & 	4.04 & 	10.7  \\ 
FB22 & 	6944 & 	0.5877 &	4.75 & 	0.028 & 	4.04 & 	10.3  \\ 
FB23 & 	6890 & 	0.5935 &	4.78 & 	0.028 & 	4.04 & 	10.3  \\ 
FB27 & 	7214 & 	0.5991 &	4.86 & 	0.030 & 	4.04 & 	10.0  \\ 
FB24 & 	7054 & 	0.6039 &	4.89 & 	0.031 & 	4.04 & 	9.3  \\ 
 \end{tabular}
 \end{ruledtabular}
 \end{table}

 \begin{table}
 \caption{\label{t.2}
Details for the dry acrylic samples in the internal region of the DEM relaxed samples:
Number of internal mobile spheres (N), Packing fraction ($\phi$), Average number of neighbors ($z_c$), Fraction of contacts that cannot hold extra tangential force ($q$), Prediction for the isostatic condition from Eq.\ref{ISO} ($z_{ISO}$), critical size ($l^*$) in units of $d$.
}
\begin{ruledtabular}
 \begin{tabular}{lllllll}
 Sample & $N$  & $\phi$      &  $z_c$       &   $q$    &  $z_{ISO}$ &  $l^*$ \\ \hline
B & 	7041 & 	0.5971 &	4.88 & 	0.150 & 	4.21 & 	11.6  \\ 
A & 	6922 & 	0.5997 &	4.90 & 	0.163 & 	4.23 & 	11.0  \\ 
C & 	6804 & 	0.6148 &	4.89 & 	0.143 & 	4.20 & 	9.8  \\ 
D & 	7242 & 	0.6258 &	5.05 & 	0.162 & 	4.23 & 	8.7  \\ 
E & 	6857 & 	0.6288 &	5.03 & 	0.152 & 	4.21 & 	9.4  \\ 
F & 	6856 & 	0.6367 &	5.05 & 	0.169 & 	4.24 & 	9.3  \\ 
 \end{tabular}
 \end{ruledtabular}
 \end{table}

 \begin{table*}
 \caption{\label{t.3}
Estimation of the neighbors in contact computed by means of the deconvolution method in Ref.\cite{AstePRE05} Table.I $z_{DE}$.
Number of near neighbors  computed from $g_z(r)$ at $r-d/2=0.004d$ and from the maximum of $g_z(r)$.
The number of actual contacts $z_c$ are also reported for comparison.
}
  \begin{tabular}{cccccccccc}
 \hline
 \hline
 Sample & $z_{DE}$  & $g_z(0.004d+d/2)$ &  $max(g_z(r))$  & $z_c$ & Sample & $z_{DE}$ & $g_z(0.004d+d/2)$  &  $max(g_z(r))$ & $z_c$ \\ \hline
FB14 &	5.60 & 4.91  &	5.29 &	4.70  & A &	5.81 & 5.26  &	5.75   &	4.90  \\ 
FB15 &	5.70 & 4.88  &	5.28 	&	4.71  & B &	5.91 & 5.25  &	5.76  &	4.88\\ 
FB16 &	5.30 & 4.88   &	5.30 	&	4.71  & C &	6.77 & 5.34  &	6.09  &	4.89\\ 
FB17 &	5.40 & 4.89  &	5.29 	&	4.69  & D &	6.78 & 5.55 &	6.33  &	5.05\\ 
FB18 &	5.00 & 4.86  &	5.25 	&	4.70  & E &	6.95 & 5.58  &	6.39  &	5.03\\ 
FB19 &	5.30 & 4.89  &	5.31 	 &	4.72  & F &	6.97 & 5.61  &	6.56  &	5.05\\ 
FB20 &	5.80 & 4.86  &	5.29 	&	4.69  \\ 
FB21 &	5.80 & 4.92  &	5.32 	&	4.74  \\ 
FB22 &	5.70 & 4.92  &	5.36 	 &	4.75  \\ 
FB23 &	5.90 & 4.96  &	5.43 	&	4.78  \\ 
FB24 &	5.90 & 5.06  &	5.63 	 &	4.89  \\ 
FB27 &	6.00 & 5.02   &	5.54 &	4.86  \\ 
\hline
\hline
 \end{tabular}
 \end{table*}

%

%

\begin{figure}
\centering
\includegraphics[width=0.49\columnwidth]{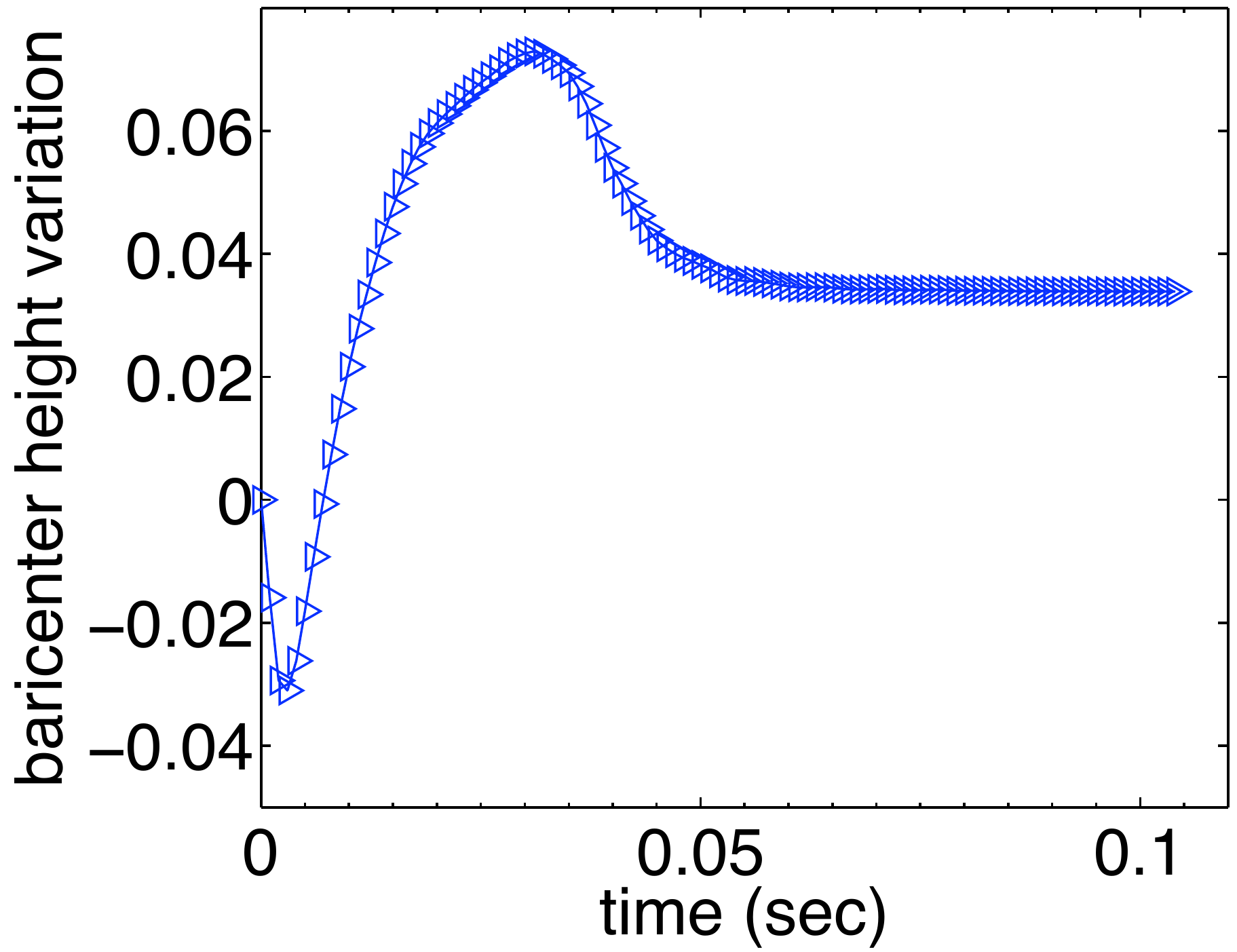}
\includegraphics[width=0.49\columnwidth]{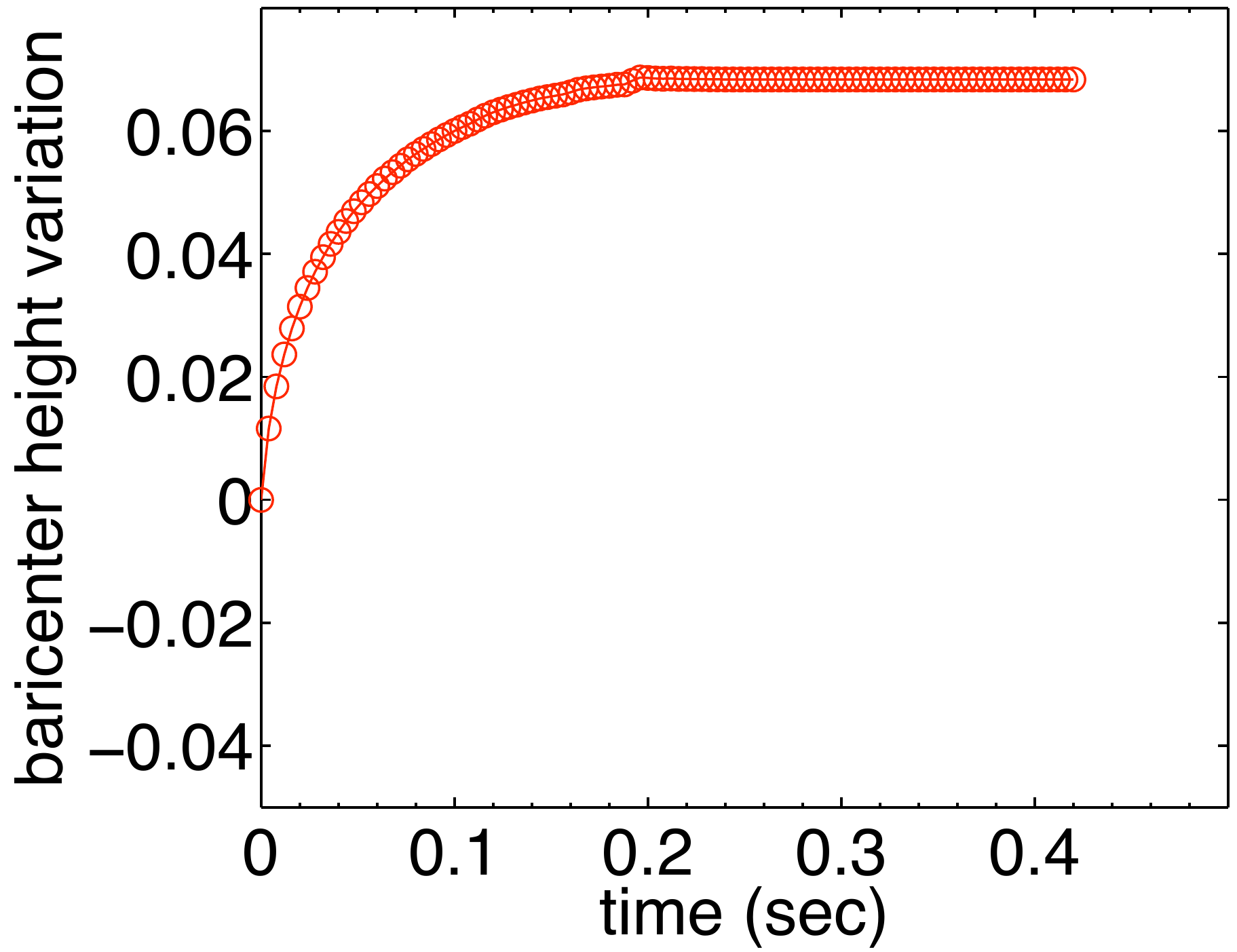}
\caption{ 
Variation of the barycenter height (in unit of $d$) during a DEM relaxation simulation. 
The left plots refer to the sample with lowest packing fraction in the fluidized bed experiments (sample FB18) and the right plot is the sample with the largest packing fraction in the dry beads experiment (sample F).
}
\label{f:height}
\end{figure}

\begin{figure}
\centering
 \begin{tabular}{cc}
 \includegraphics[width=0.5\columnwidth]{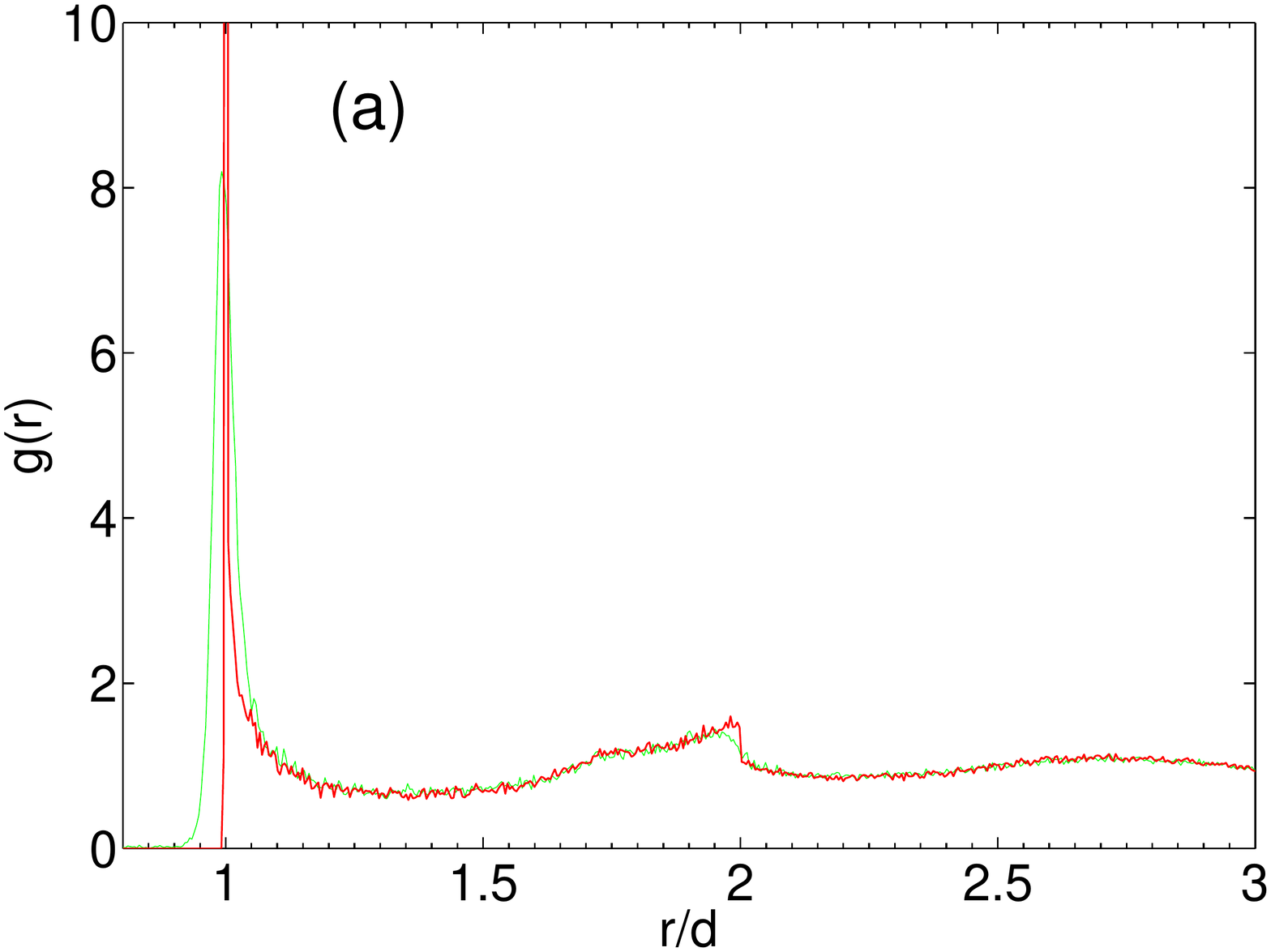}&
\includegraphics[width=0.5\columnwidth]{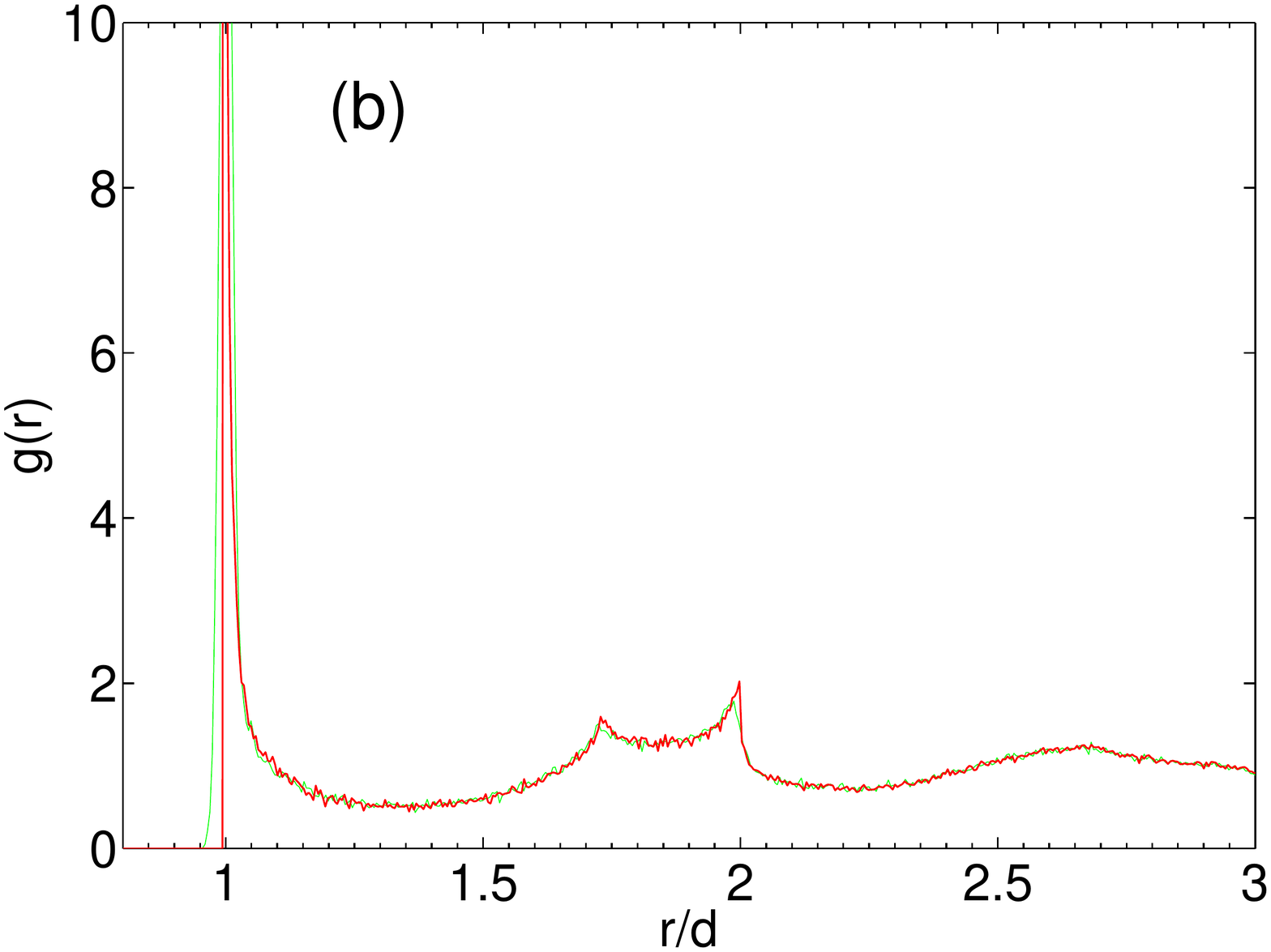}\\
\includegraphics[width=0.5\columnwidth]{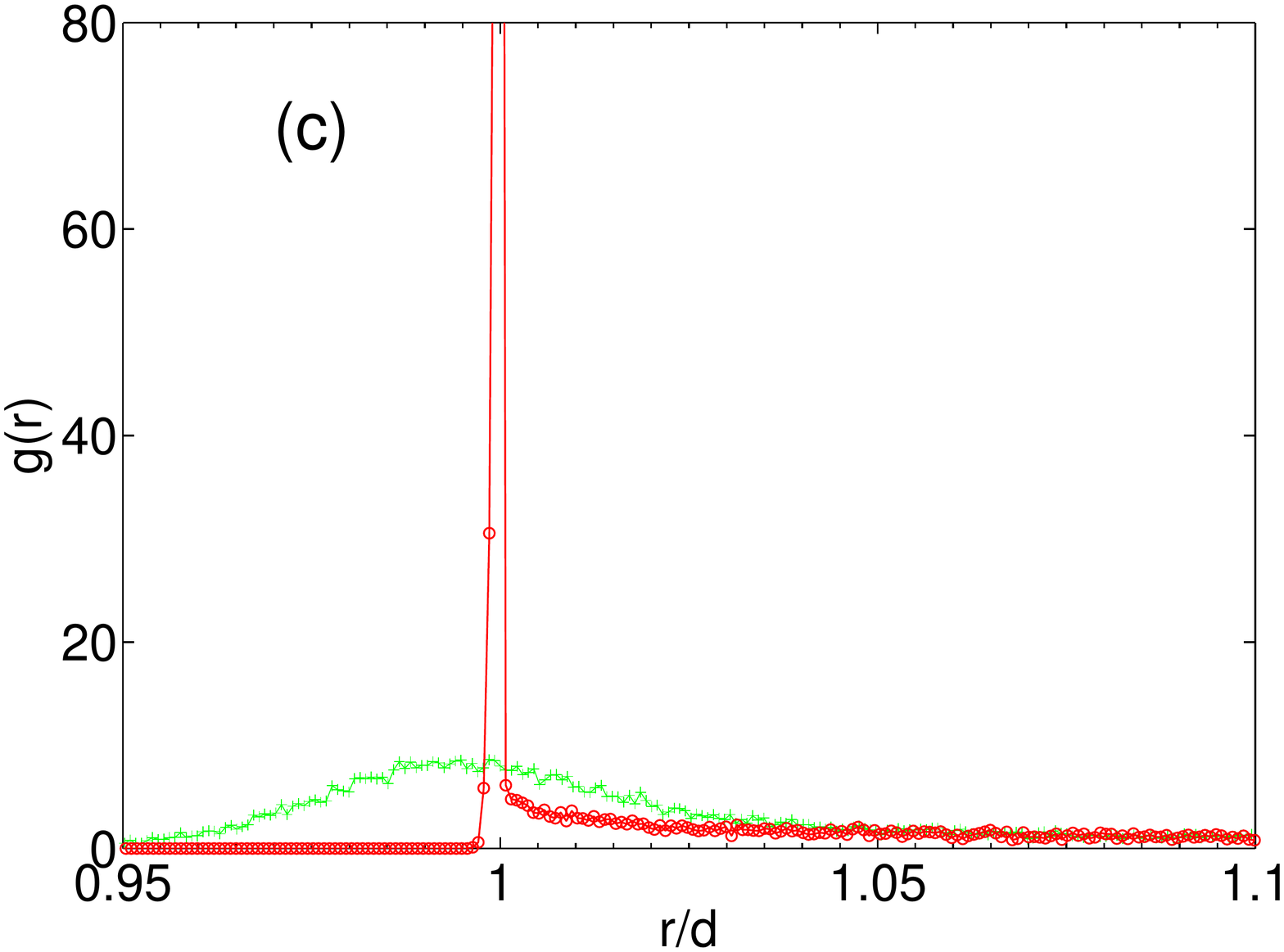}&
\includegraphics[width=0.5\columnwidth]{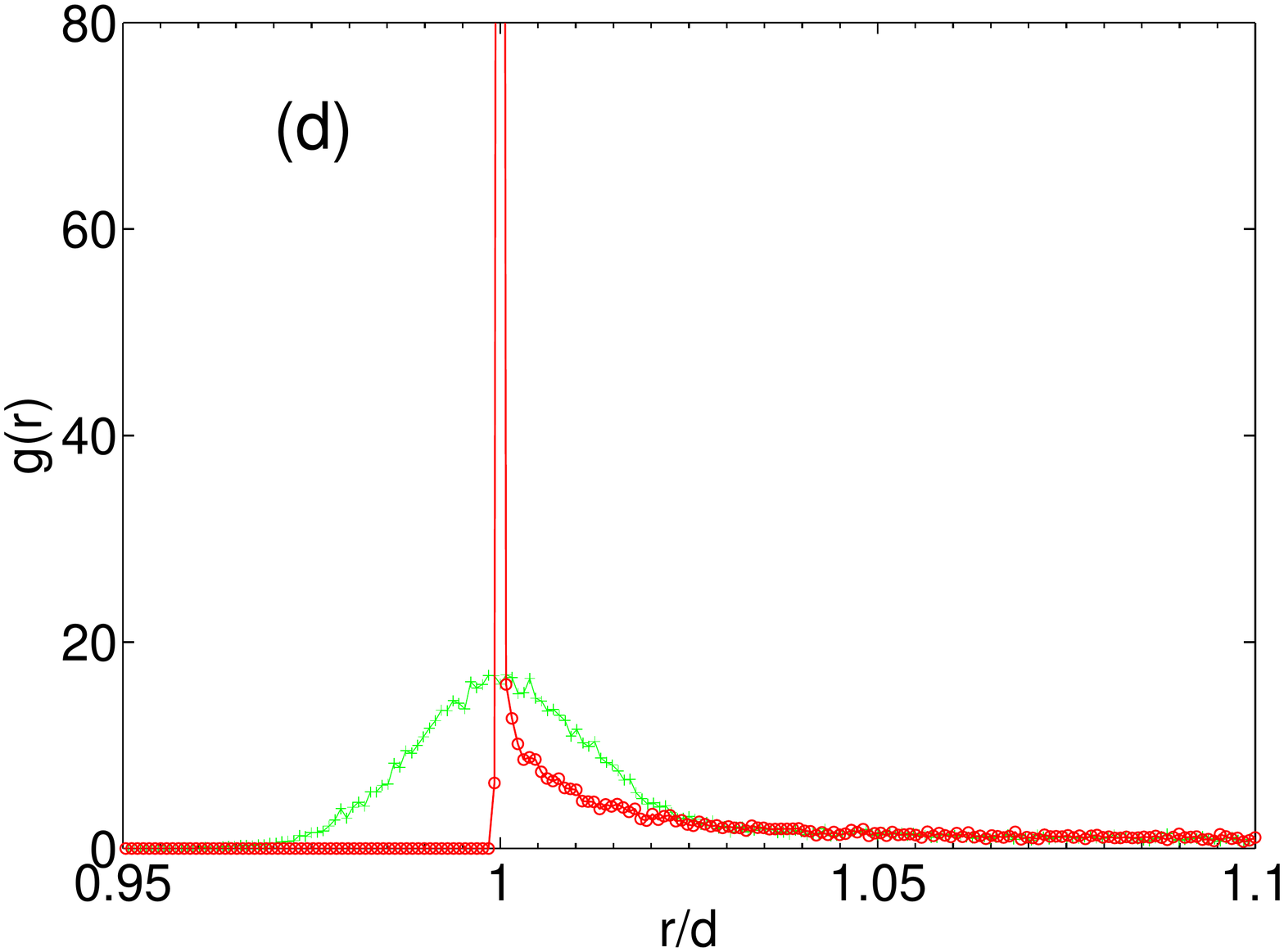}\\
\includegraphics[width=0.5\columnwidth]{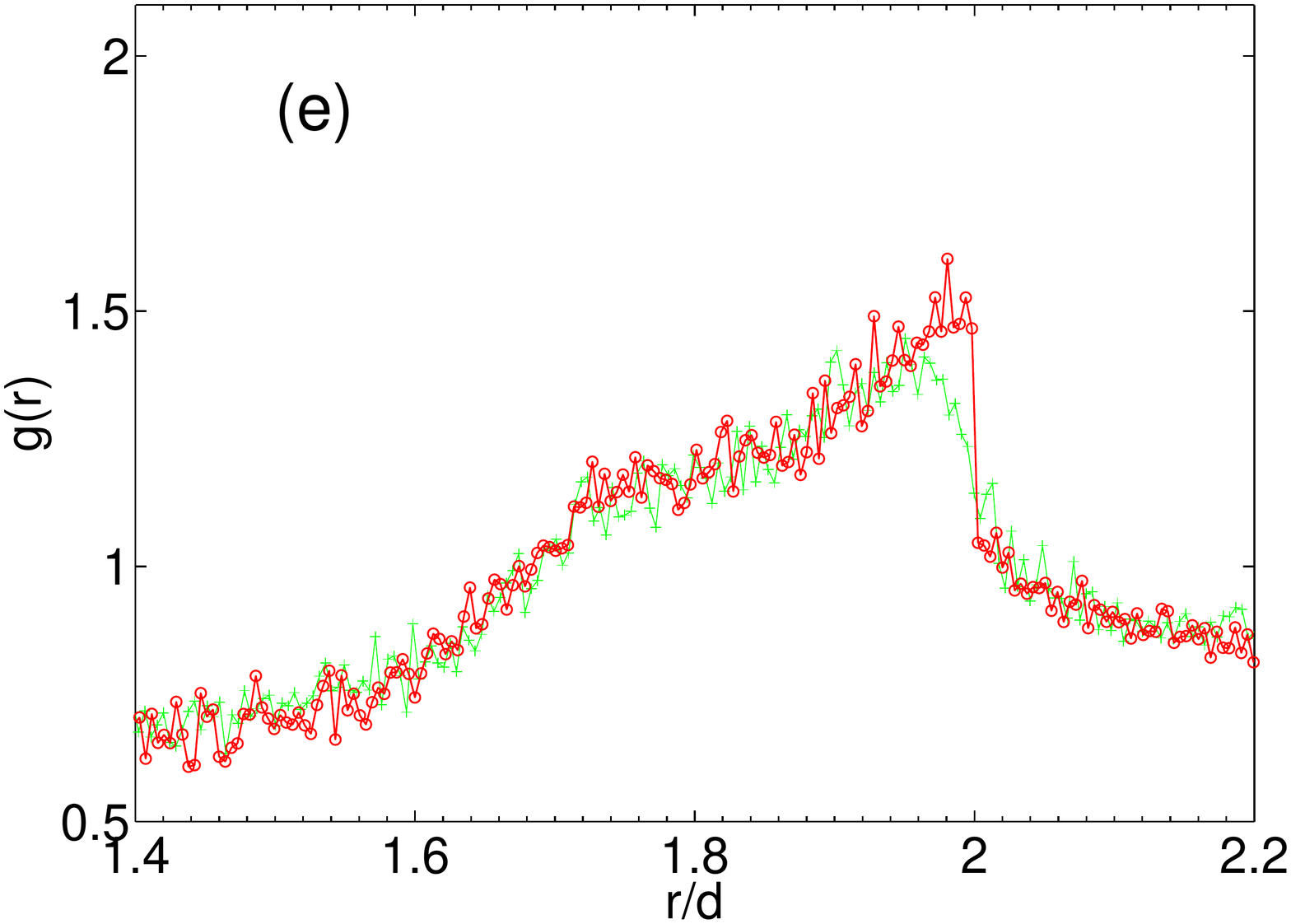}&
\includegraphics[width=0.5\columnwidth]{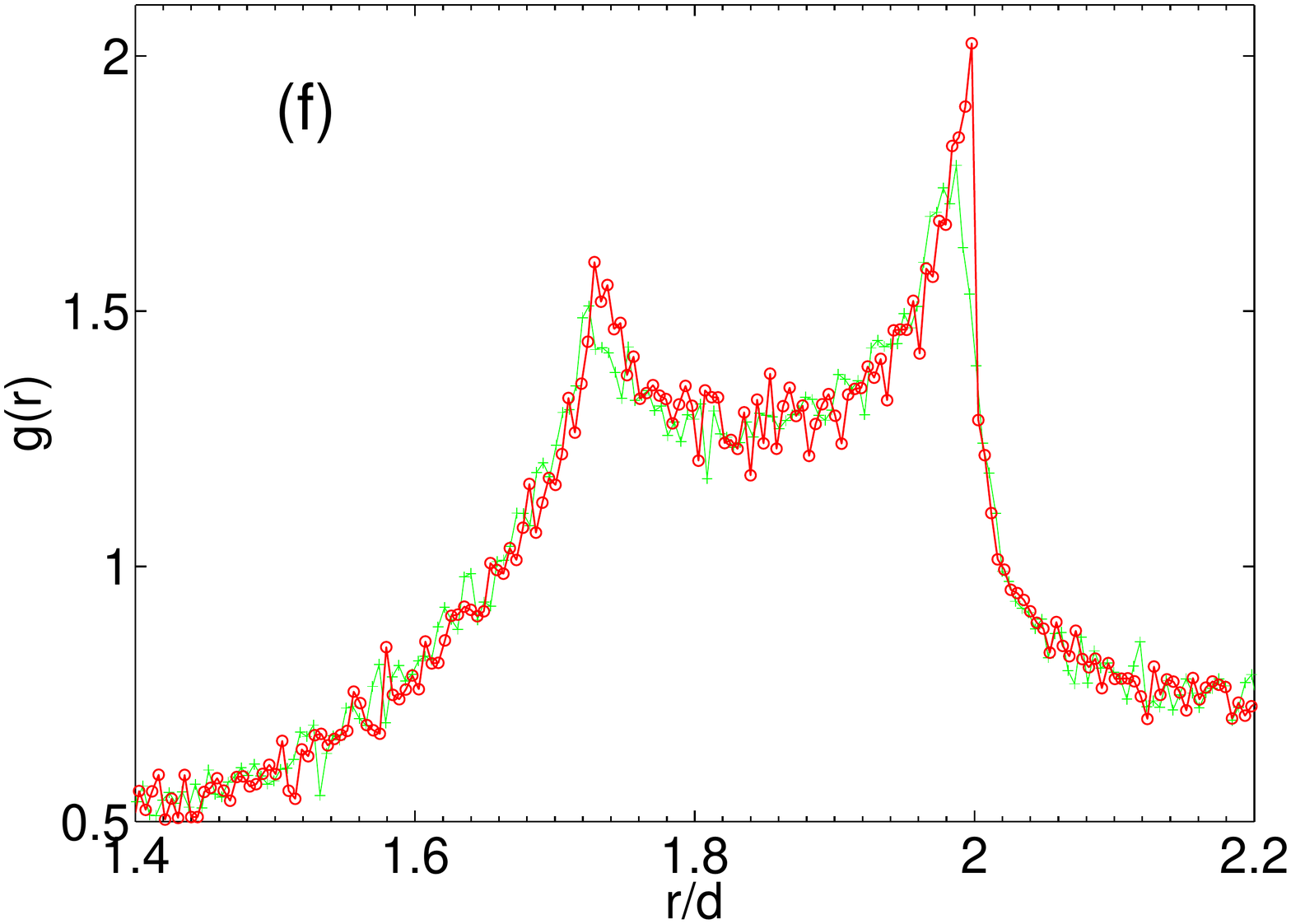}\\
\end{tabular}
\caption{  
Comparison between the radial distribution functions from the original tomographic data and from the DEM relaxed.
The data refer respectively to (left plots: a,c,e) the sample with lowest packing fraction in the fluidized bed experiments (sample FB18) and (right plots: b,d,f) the sample with the largest packing fraction in the dry beads experiment (sample F).
}
\label{f.gdr}
\end{figure}

\section{Radial Distribution Function}

\begin{figure}
\centering
 \begin{tabular}{cc}
\includegraphics[width=0.5\columnwidth]{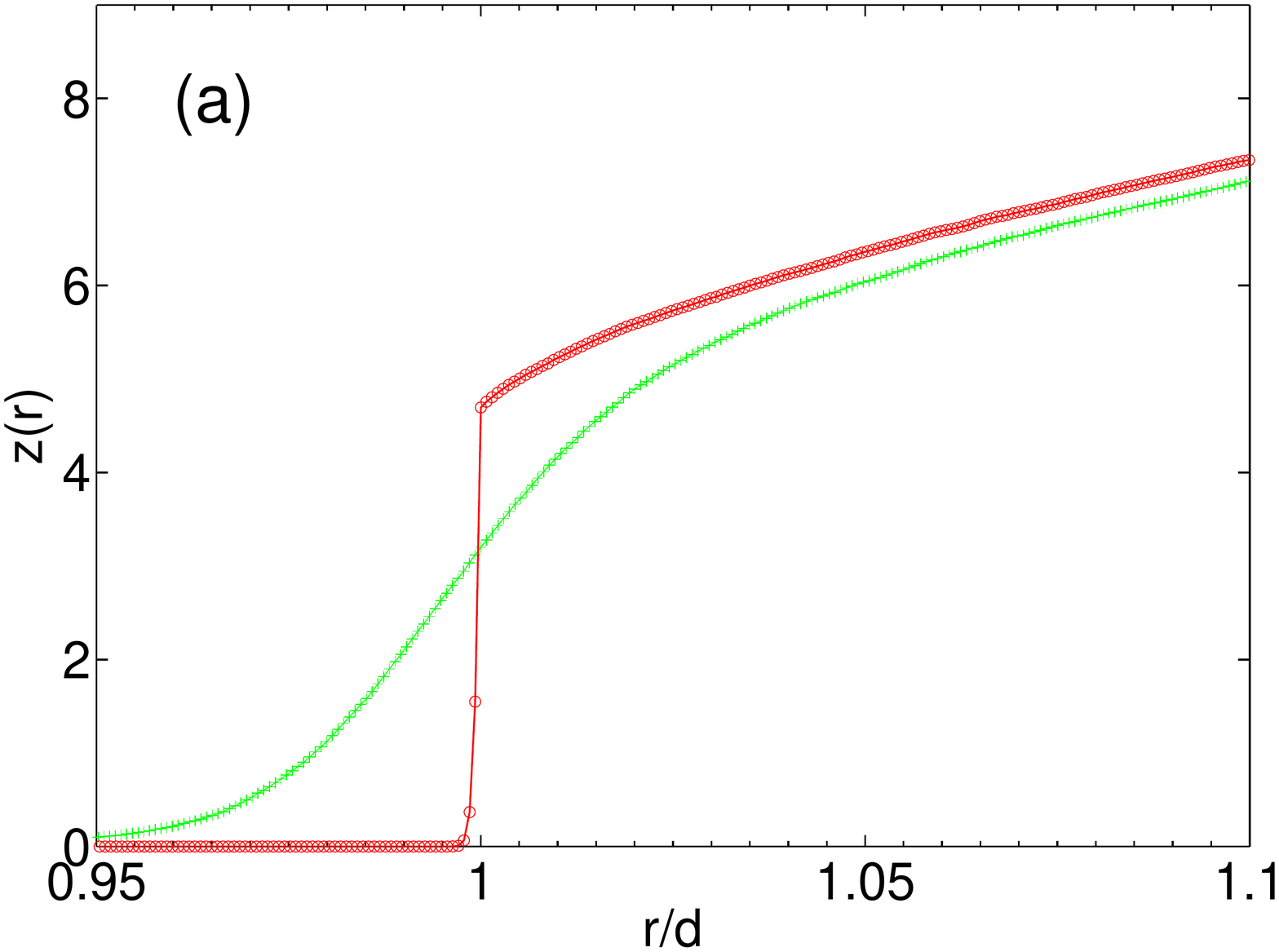}&
\includegraphics[width=0.5\columnwidth]{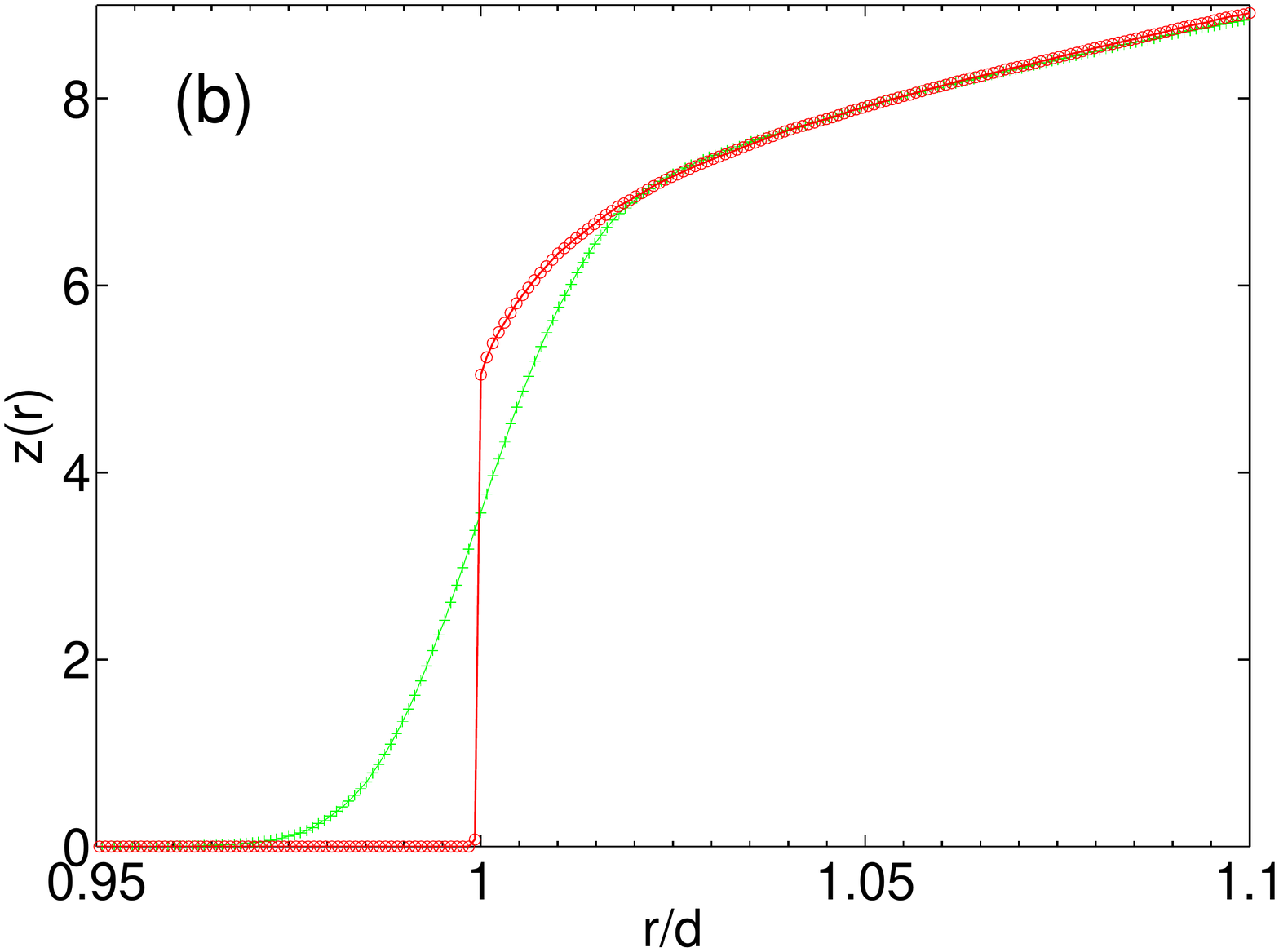}\\
\end{tabular}
\caption{ 
Average number of sphere centers within a distance $r$ form the centre of a sphere.
The DEM relaxed samples (red circles) reveal a sharp discontinuity at $r=d$.
The left plot refer to the sample with lowest packing fraction in the fluidized bed experiments (sample FB18) and right to the sample with the largest packing fraction in the dry beads experiment (sample F). 
}
\label{f:cdf}
\end{figure}

A classical quantity widely used to characterize the structural organization of packings is the the radial distribution function $g(r)$ which is associated to the probability of finding the center of a sphere in a given position at distance $r$ from a reference sphere. 
This quantity is calculated by counting the number of sphere centers within a radial distance $r$ from a given sphere center ($z(r)$, see Fig.\ref{f:cdf}) and using
\begin{equation}
z(r_1)-z(r_0) = \int_{r_0}^{r_1} g(r) 4 \pi r^2 \, dr.
\label{zg}
\end{equation} 
Although there are more specific better devised methods to investigate and characterize the  geometrical organization in disordered packings \cite{Aste05rev}, the comparison between $g(r)$ from the original and the DEM relaxed samples is a very good way to measure average variation in the internal structure.
Figure \ref{f.gdr} shows the radial distribution function $\tilde{g}(r)$ (which has been normalised such that $\tilde{g}(r) \rightarrow 1$ for $r \rightarrow \infty$) for both the DEM relaxed samples and and the original tomographically obtained data. 
We observe that both sets of data show a peak at $r = d$, which is however much sharper in the DEM simulation. 
We also observe two peaks at $r = \sqrt{3}d$ and $r \simeq 2d$, which are again sharper in the DEM simulation but they preserve the overall characteristics of the experimental data.
This is a strong indication that there are essentially no differences between geometrical properties in the original experimental samples and in the DEM relaxed samples except for the acquired uniformity of the sphere diameters.

\begin{figure}
\centering
\includegraphics[width=0.49\columnwidth]{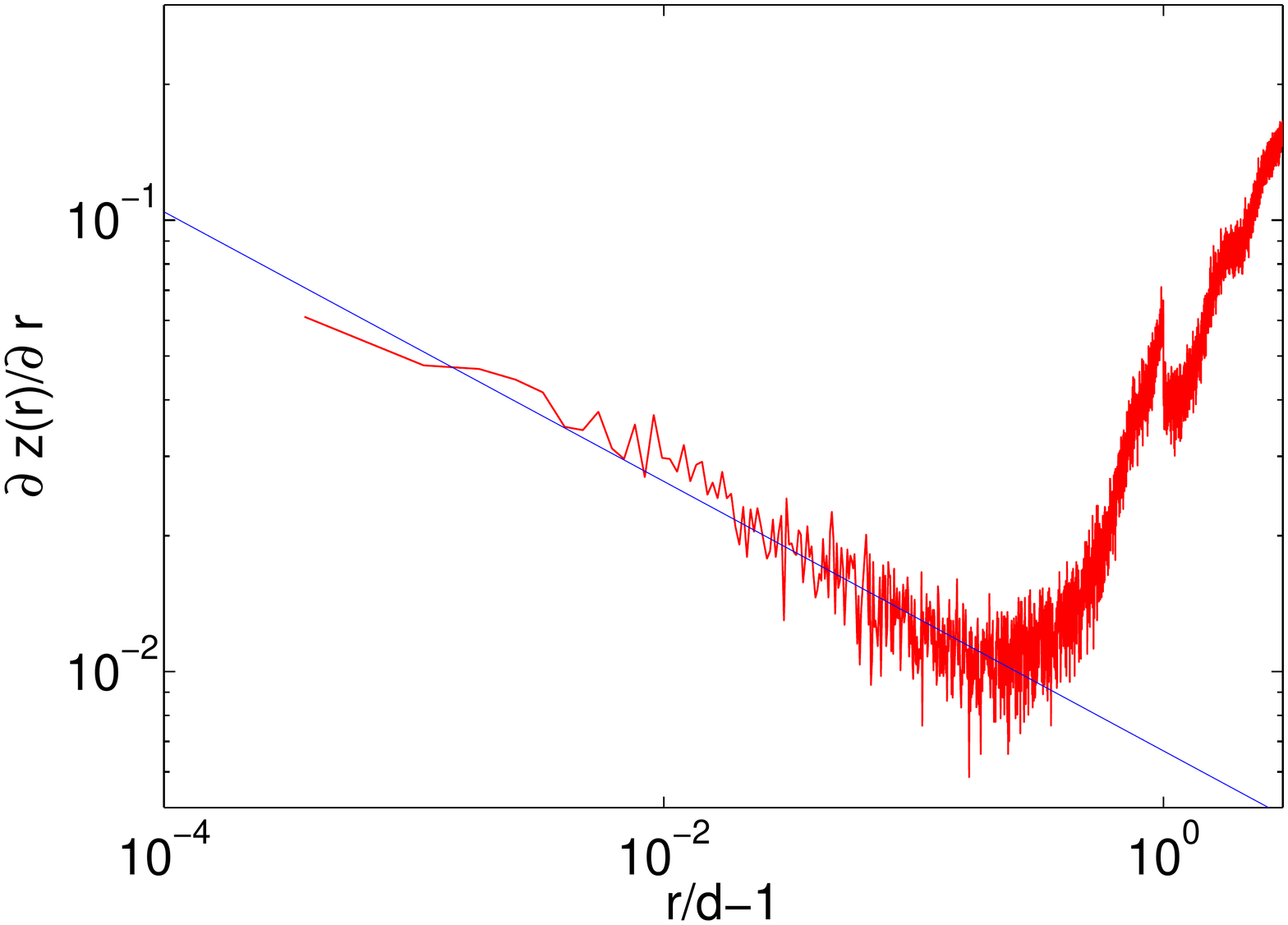} 
\includegraphics[width=0.49\columnwidth]{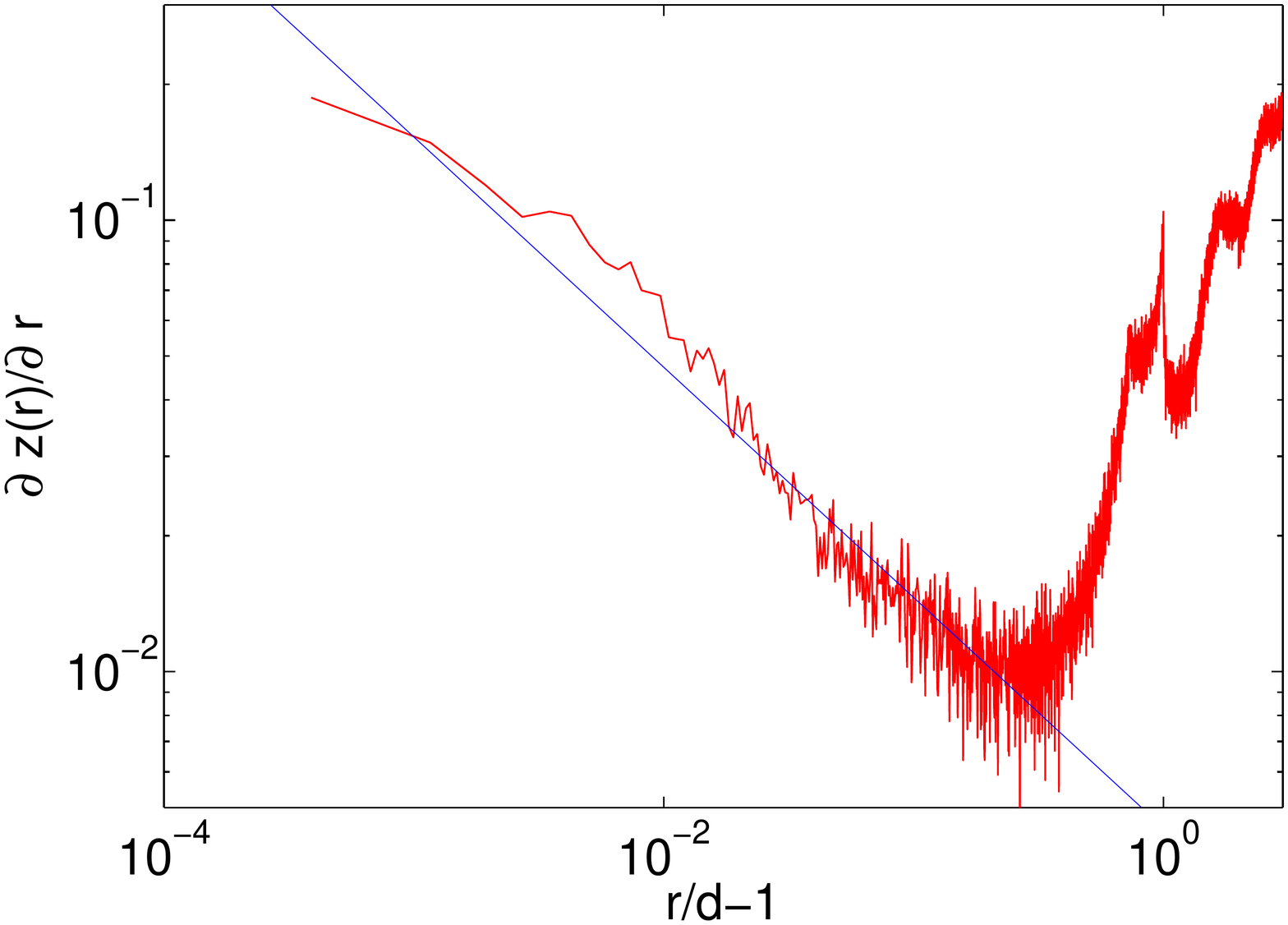}
\caption{ 
The rate of variation of the number of neighbours vs. radial distance shows power law behaviors: $\partial z(r)/\partial r = (r-d)^{\alpha-1}$ (liner trend in log-log scale in the region $r\in [d,1.4d]$).
(top) Fluidized bed sample FB18; (bottom) dry acrylic sample F.
The exponent depend on sample preparation and packing fraction. 
The lines are the best fits in the region $r \in [d,1.4d]$.
}
\label{f2000}
\end{figure}

\begin{figure}
\centering
\includegraphics[width=1.0\columnwidth]{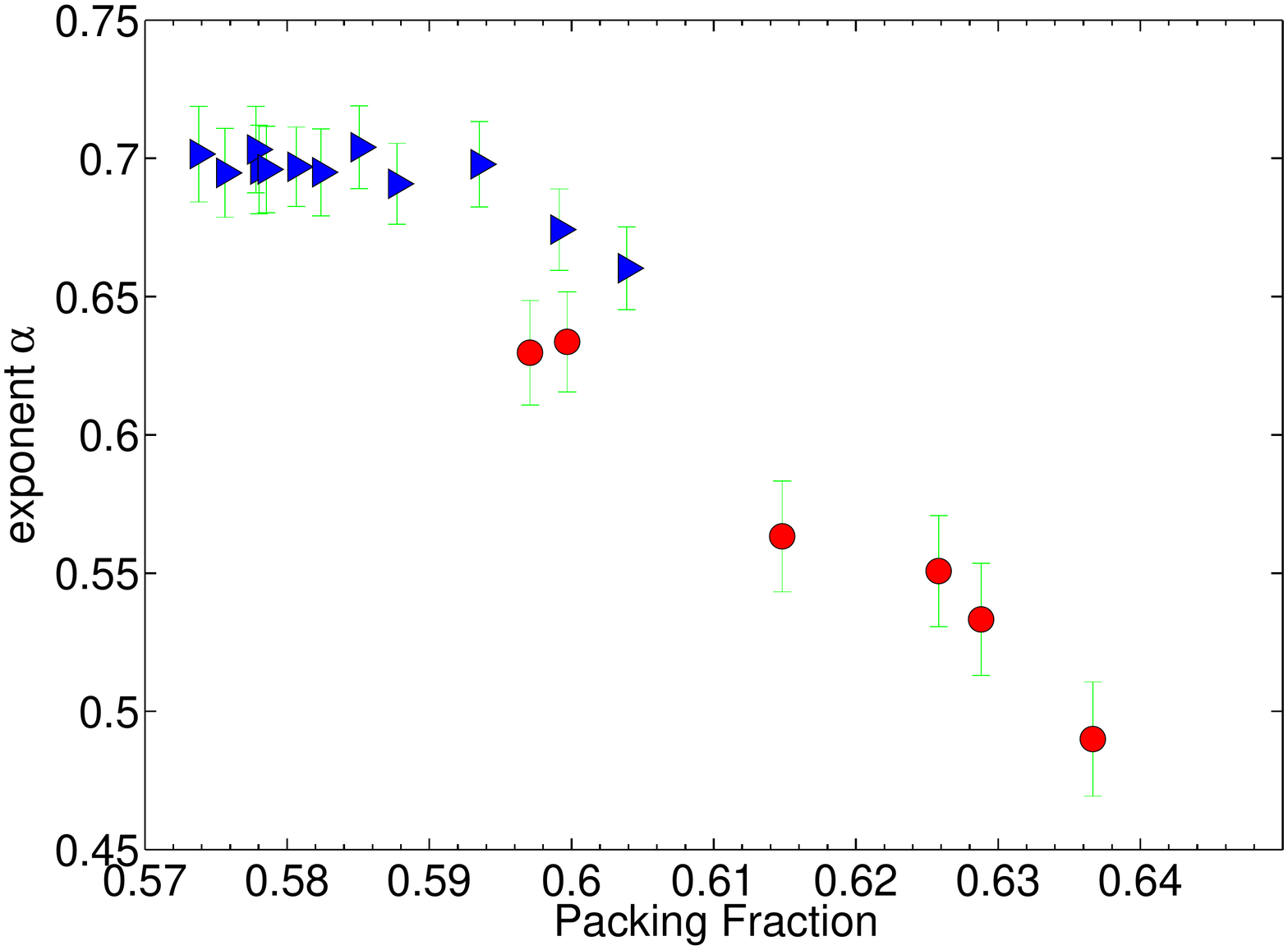} \\
\caption{ 
Exponent $\alpha$ vs. packing fraction for all the samples.
The error bars represent 95\% confidence interval over the estimation of $\alpha$ from linear regression.
}
\label{f2001}
\end{figure}

\subsection{Power law increase in the number of neighbors}

The average number of neighbors $z(r)$ that stay within a given radial distance $r$ from a sphere centre in the packing is a quantity directly related to the $g(r)$ through Eq.\ref{zg}.
For a packing of perfect, non-overlapping spheres $z(r)$ must be identically equal to zero for any $r<d$ and then it must jump to $z_c$ at $r=d$.
In Fig.\ref{f:cdf} we can see that $z(r)$ from the experimental samples has a rather smooth increase  with value different from zero in a relatively large range of distances up to 5\% below $r=d$.
A very different behavior is revealed by the DEM relaxed samples where a jump to $z_c$ is clearly visible at $r=d$.

For larger distances, $z(r)$ keeps growing gently while other  centers from nearby spheres that are not in contact become within reach.
It has been pointed out in the literature \cite{Donev05a,Kamien07} that, after the jump at $r=d$, the number of neighbors within a given radial distance $z(r)$ increases smoothly with the distance $r$ accordingly with the law (for $r>d$):
\begin{equation}
z(r) = z_c + z_1(r-d)^\alpha \;\;;
\label{powLaw1}
\end{equation}
which implies 
\begin{equation}
\frac{\partial z(r)}{\partial r} = z_1 \alpha (r-d)^{\alpha-1} \;\;.
\label{powLaw2}
\end{equation}
It has been argued that such a power law behavior has interesting relations with the statistical physics of these systems and it is related with the dynamical arrest at jamming transition  \cite{Ojha04,Silbert06,Kamien07}.

Figure \ref{f2000} reports $\partial z(r)/\partial r$ vs. $r$ for the two samples with respectively the smallest and largest packing fractions (FB18 and F).
As one can see both the samples show a good linear trend in log-log scale revealing that the law in Eq.\ref{powLaw2} is well followed. 
Similar behaviors are observed for all samples.
However, it turns out that the exponent $\alpha$ are not universal.
Indeed, they change with  packing fraction and they depend on the sample preparation as shown in Fig\ref{f2001}.
This is in apparent contradiction with the results and discussion in  \cite{Donev05a,Kamien07} which suggest a `universal'  value of 0.5 for the exponent.
However, it must be stressed that numerical results in \cite{Donev05a,Kamien07} refer to non-frictional spheres that in a sedimentation experiment are only stable at the limiting packing fraction of $0.64$.
From Fig\ref{f2001} one can see that our data are consistent with $\alpha = 0.5$ at $\Phi = 0.64$.

\begin{figure}
\centering
\includegraphics[width=1.0\columnwidth]{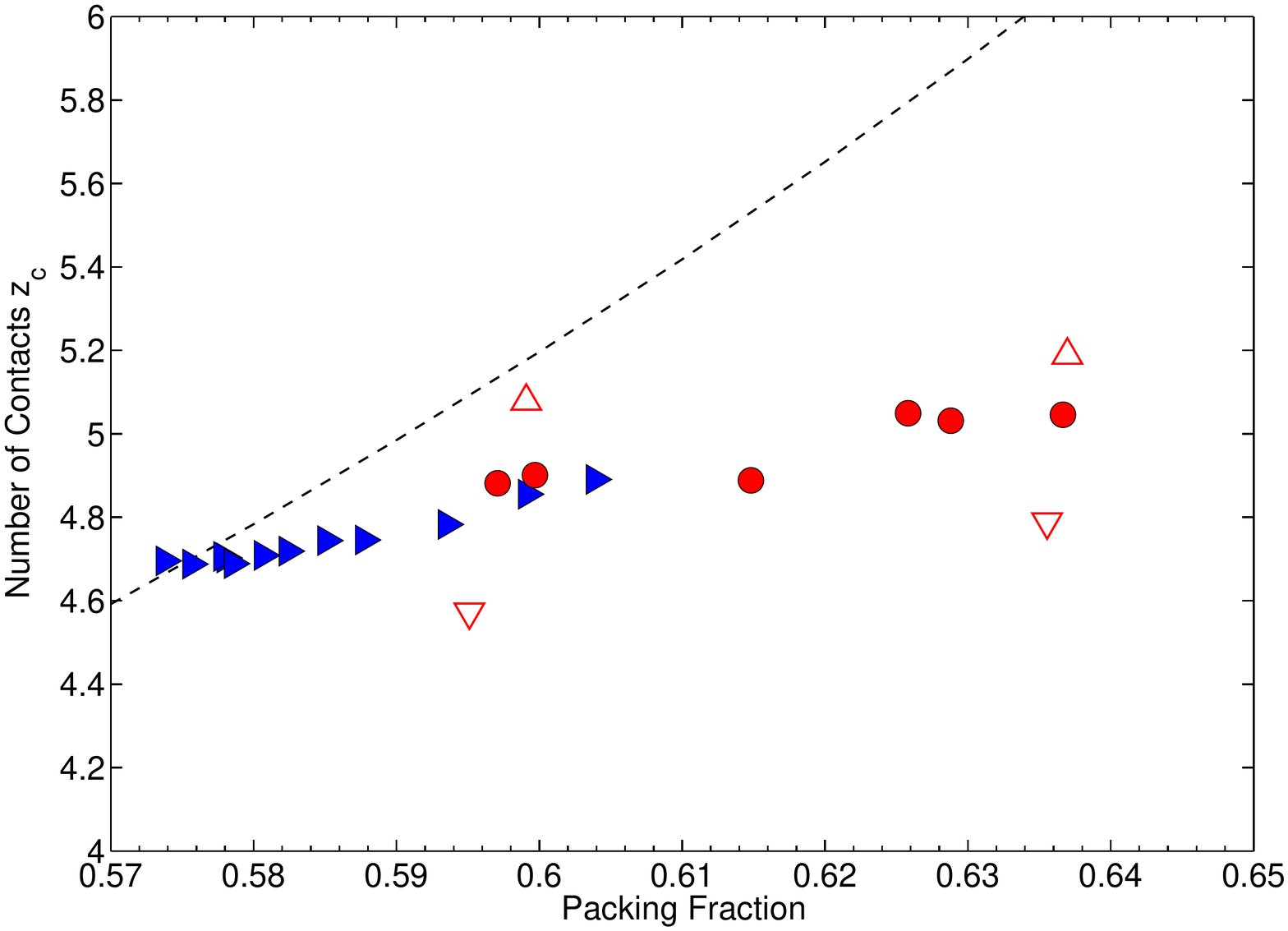}
\caption{  
Average number of neighbors in contact vs. packing fraction for all the experimental samples: fluidized bed ($\triangleright$) and dry acrylic beads ($\circ$).
The dashed line is the theoretical prediction from \cite{Song08} $z_c = 2\sqrt{3} \Phi/(1-\Phi)$.
The $\bigtriangledown$ refer to two samples (B and F) relaxed with a larger friction coefficient ($\mu =0.9$ instead of 0.28).
Conversely, the $\bigtriangleup$ refer to samples B and F relaxed with a smaller friction coefficient ($\mu = 0.2$).
}
\label{f.contactsDec}
\end{figure}

\begin{figure}
\centering
\includegraphics[width=.49\columnwidth]{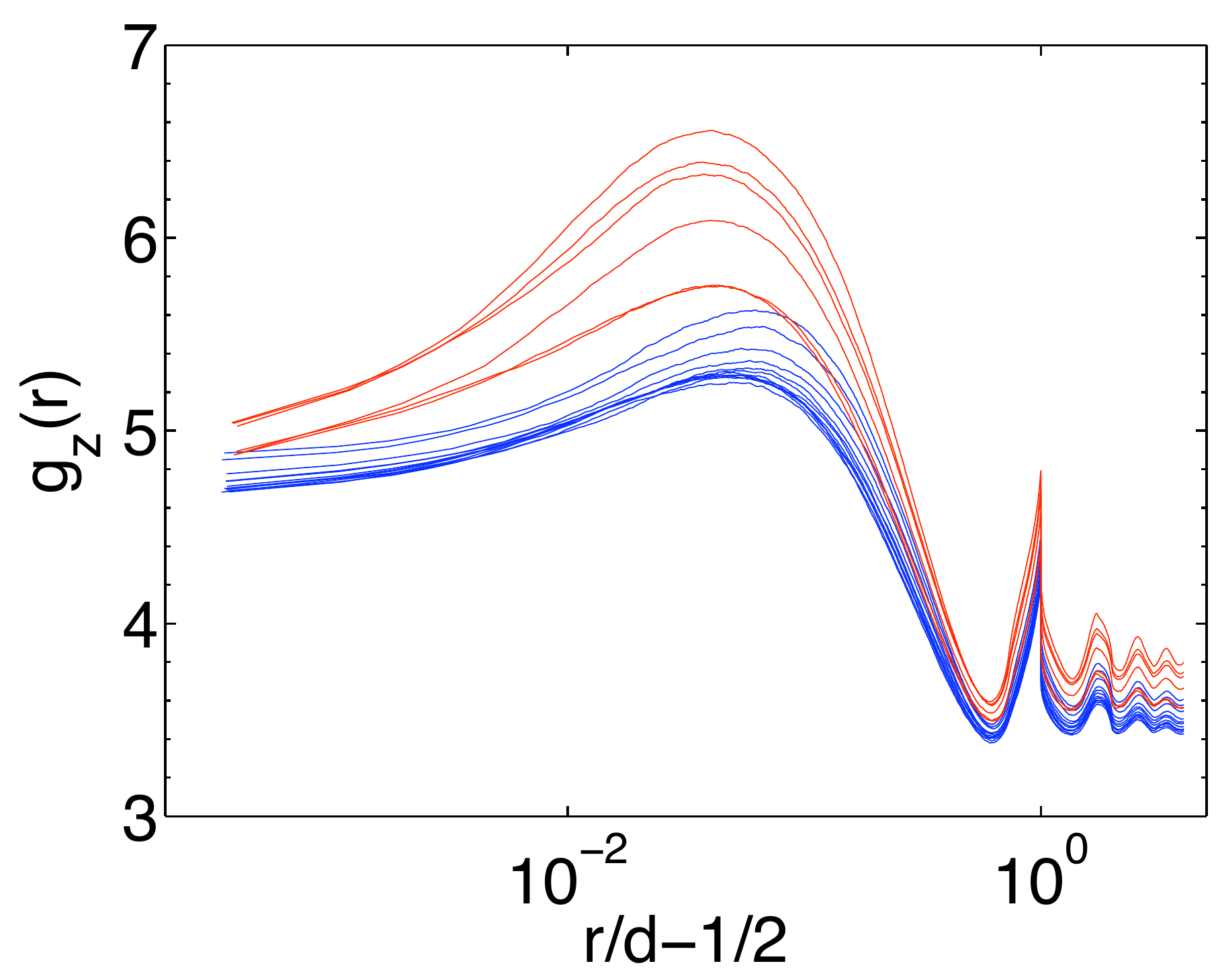}
\includegraphics[width=.49\columnwidth]{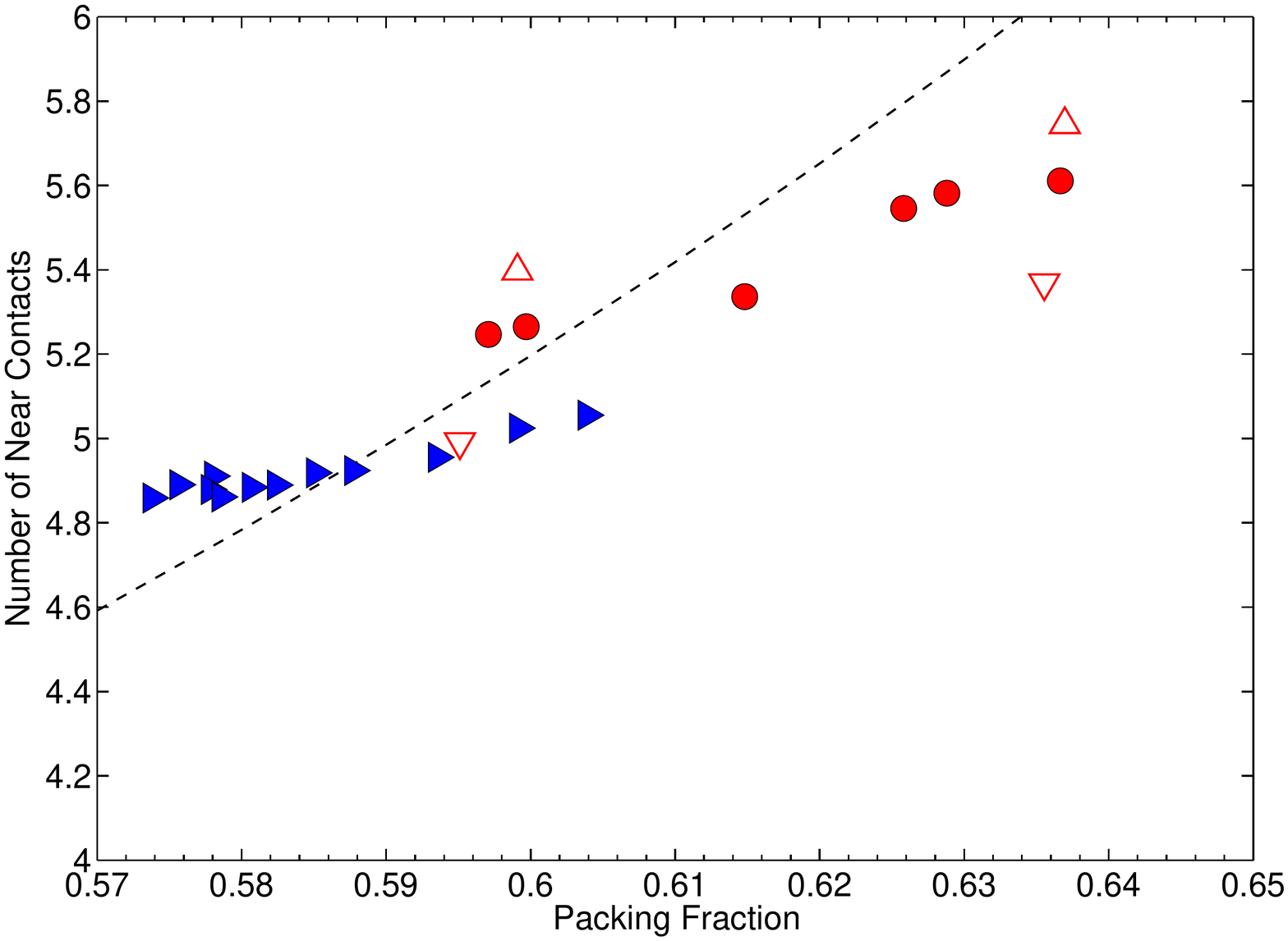}\\
\caption{ 
(top) 
Modified radial distribution function $g_z(r)$ (see text).
(bottom) 
Number of near contacts estimated from the value of $g_z(r)$ at $r-d/2 = 0.004d$.
The dashed line is  $z = 2\sqrt{3} \Phi/(1-\Phi)$ \cite{Song08}.
}
\label{g_z_x_Macse}
\end{figure}

\section{Number of contacts}

The average number of neighbors in contact with any given sphere in the packing ($z_c$)  is a simple and important measure of the system's topological structure and it has recently become central in the theoretical description of these systems by means of statistical mechanics arguments \cite{Song08}.
The estimation of this quantity from geometrical structural measurements is extremely problematic because grains can be infinitesimally closed but not in mechanical contact.
This problem can only be solved if one can measure the amount of stress between grains. 
Direct observation of the stress with photo-realistic materials is possible but it is extremely challenging in three dimensions and to our knowledge there is still are no available data. 
On the other hand, with tomographic data we cannot measure directly the forces between grains, however by DEM relaxing the systems we can.
Indeed, one of the main initial motivations for the present study was to provide a precise measurement of the number of grains in contact.

Figure \ref{f:cdf} shows the comparison between the behavior of the average number of sphere centers $z(r)$ within a radial distance $r$ obtained from both  the DEM relaxed samples and the original tomographic data. 
It is clear that the determination of the average number of neighbors in contact from the original experimental data is extremely difficult task because the data have a slow smooth increase in the number of neighbours over a range of $r = 0.97d \rightarrow 1.02d$. 
Conversely, the DEM relaxed samples  show a much sharper increase in the number of neighbors, with a discontinuity at $r \simeq d$ giving a good estimate of the actual average number of contacts $z_c$. 
Such an estimate of the number of contacts for all the DEM relaxed samples  is shown in Fig.\ref{f.contactsDec}.
We observe a linearly increasing trend in the number of contact with the packing fraction with the two extreme being  $z_c \sim 4.7$ at $\phi \sim 0.57$ to $z_c \sim 5.0$ at $\phi \sim 0.64$.
We also observe little dependence on the preparation method with the fluidized beds and dry beads samples having comparable values at the same packing fractions.
Tables \ref{t.1} and \ref{t.2} report the values of $z_c$ and $\phi$.

\begin{figure}
\centering
 \begin{tabular}{cc}
\includegraphics[width=0.5\columnwidth]{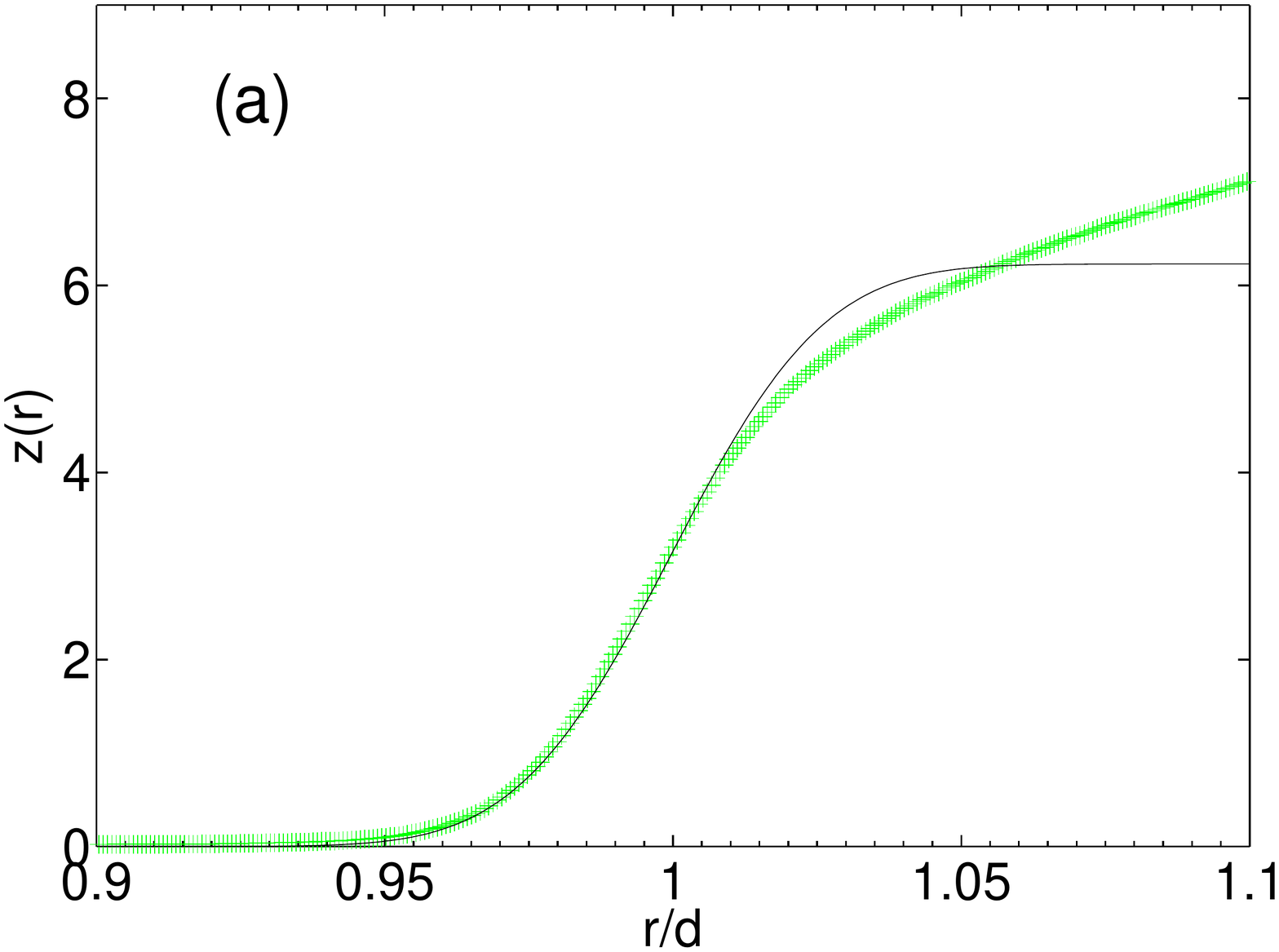}&
\includegraphics[width=0.5\columnwidth]{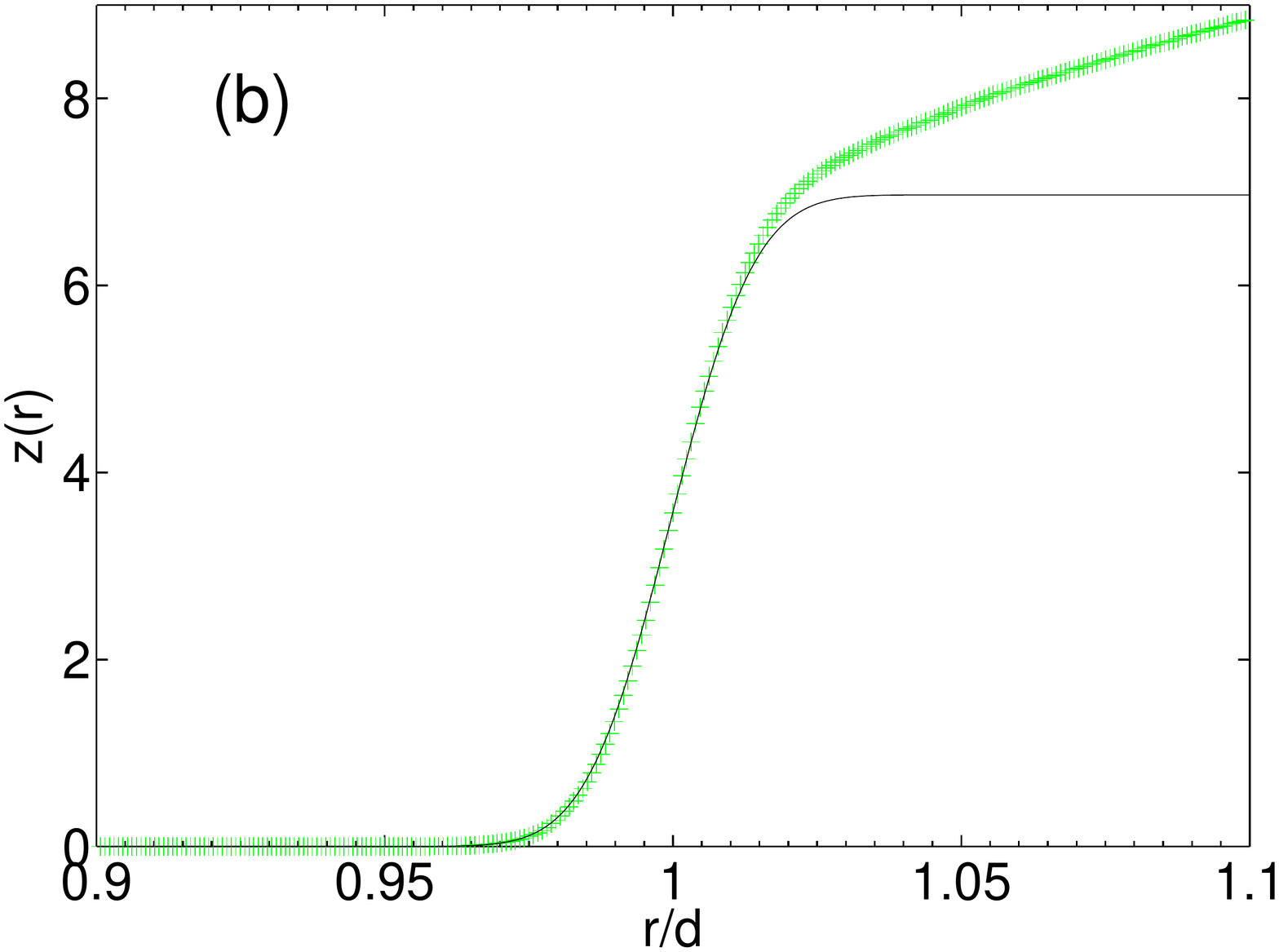}\\
\includegraphics[width=0.5\columnwidth]{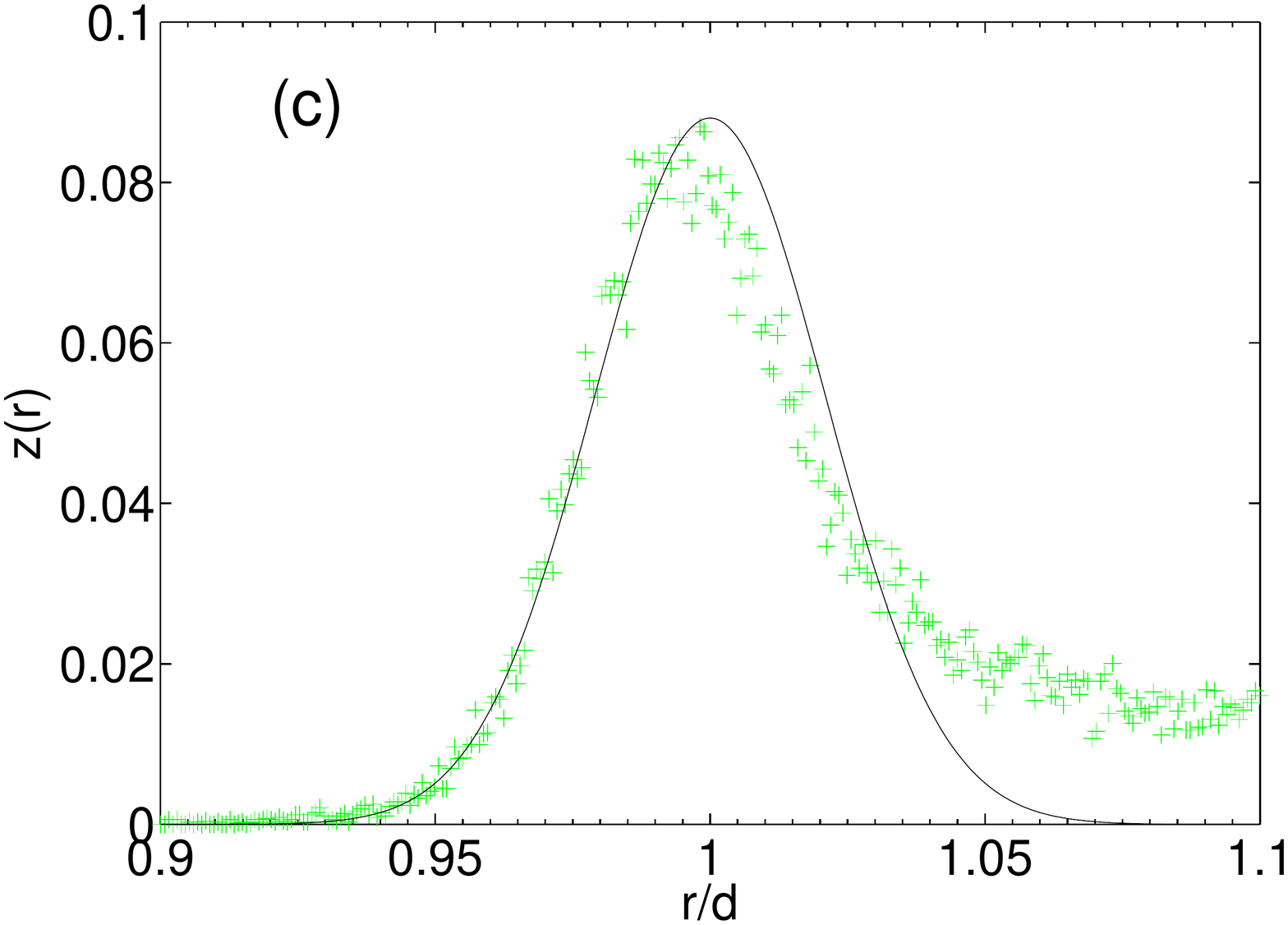}&
\includegraphics[width=0.5\columnwidth]{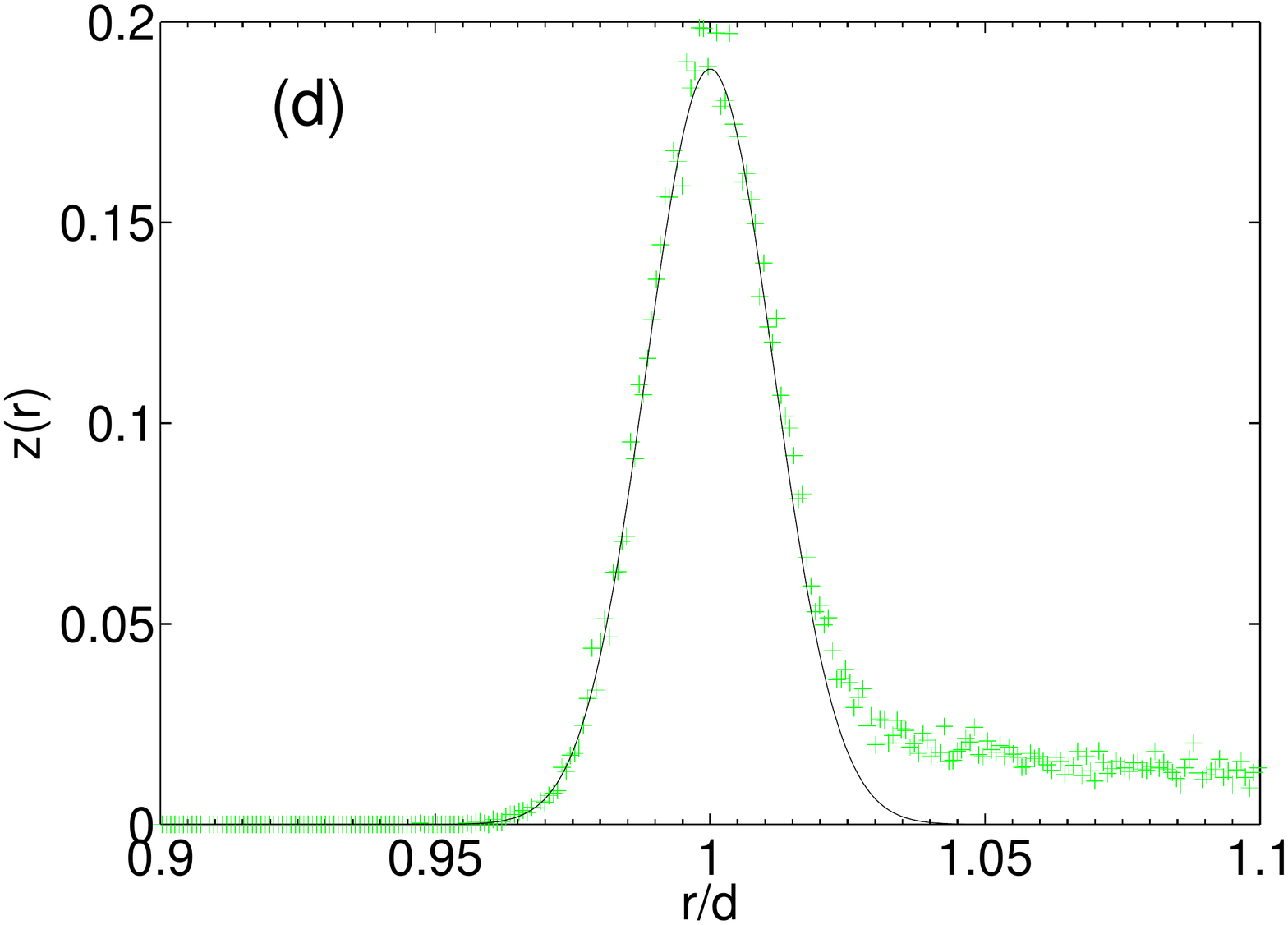}\\
\end{tabular}
\caption{ 
Two examples for samples FB18 and F of the estimation of the number of spheres in contact by means of the deconvolution method. 
The lines refers to the error functions (a,b) and the normal distributions that are best fitting the data for $r<d$. 
}
\label{f:cdf1}
\end{figure}

\subsection{Number of near neighbors}

Beside the spheres in contact there is a large number of other spheres that are near but do not touch.
These `near neighbors' certainly play a significant role in the formation of the actual packing configuration and provide the necessary environment to guarantee mechanical stability upon small perturbations.
There is however some arbitrariness in counting such neighbors. 
Indeed, as clearly visible form Fig.\ref{f:cdf}, there is a steep increase in the number of sphere centers with radial distances immediately above $r=d$. 

An interesting perspective on the analysis of the number and the role of near neighbors has been recently proposed in \ref{Song08} by using a modified radial distribution function which counts the average value of the number of grains in contact with a spherical surface radius $r$ renormalized by the factor $(d/2)^2/r^2$ which is the ratio between the surface area of the a bead and the one the spherical surface.
In our notation such a quantity is (for $r\ge d/2$):
\begin{equation}
g_z(r) = \frac{d^2}{4r^2}\left[ z(r+d/2) - z(r-d/2) \right]\;\;.
\label{Macse_gzr}
\end{equation} 
At $r=d/2$ this quantity coincides with $z_c=z(d)$, but then it has the peculiar property of showing a maximum at a radial distance which is very near (within a fraction of percent) to the contact point. 
Such a behavior is shown in Fig.\ref{g_z_x_Macse}.
From a mathematical perspective such a maximum is the consequence of the power law growth of $z(r)$  (see Eq.\ref{powLaw1}) which combined with the factor $1/r^2$ gives first a growing trend near $r=d/2$ and then a decreasing behavior at larger $r$. 
The location of such a maximum depends on the parameters $z_c$, $z_1$ and $\alpha$, but with the typical values it is within 0.5\%  from $r=d/2$.
From Fig.\ref{g_z_x_Macse} it is clear that in our samples the position of the maximum of the $g_z(r)$ varies in a relatively narrow range between $r-d/2=0.0035$ to  $r-d/2=0.0065$.

It has been argued in Ref.\cite{Song08} that a good estimation of the neighbors in near contact is to compute $g_z(r)$ at $r-d/2 = 0.004d$. 
In Fig.\ref{g_z_x_Macse} this estimation of the number of near contacts is shown as function of the sample packing fraction.
In Tab.\ref{t.3} the values are reported together with the maximal values of $g_z(r)$ in the proximity of $r/2$.

Generally speaking, it seems that beside the number of actual contacts $z_c$ which is well defined, there are isn't any clear instrument to unequivocally individuate the relevant near neighbors.

\subsection{Comparison with deconvolution method}

In a previous paper \cite{AstePRE05} the average number of grains in contact was estimated by by means of a deconvolution method using an error function to fit $z(r)$ in the region $r<d$ and assuming that the observed smooth increase is due to the grain polydispersity.
Figure \ref{f.contactsDec} shows that this method captures well the typical distribution of distances in the region $r<d$.
However, form Table \ref{t.3}, where the values of the number of contacts per sphere computed by means of this deconvolution method ($z_{DE}$) are compared with $z_c$,  it is evident that this method largely overestimates the number of actual contacts.
This is probably due to the fact that in presence of polydispersity grains that are not in contact can stay at a relative distance smaller than $d$.
Interestingly the numbers retrieved in \cite{AstePRE05} were in good agreement with several experimental estimations \cite{Bernal60,Scott62} pointing out the need for better experimental methods to directly access this quantity.

\section{Local relation between packing fraction and number of contacts}

\begin{figure}
\centering
\includegraphics[width=.49\columnwidth]{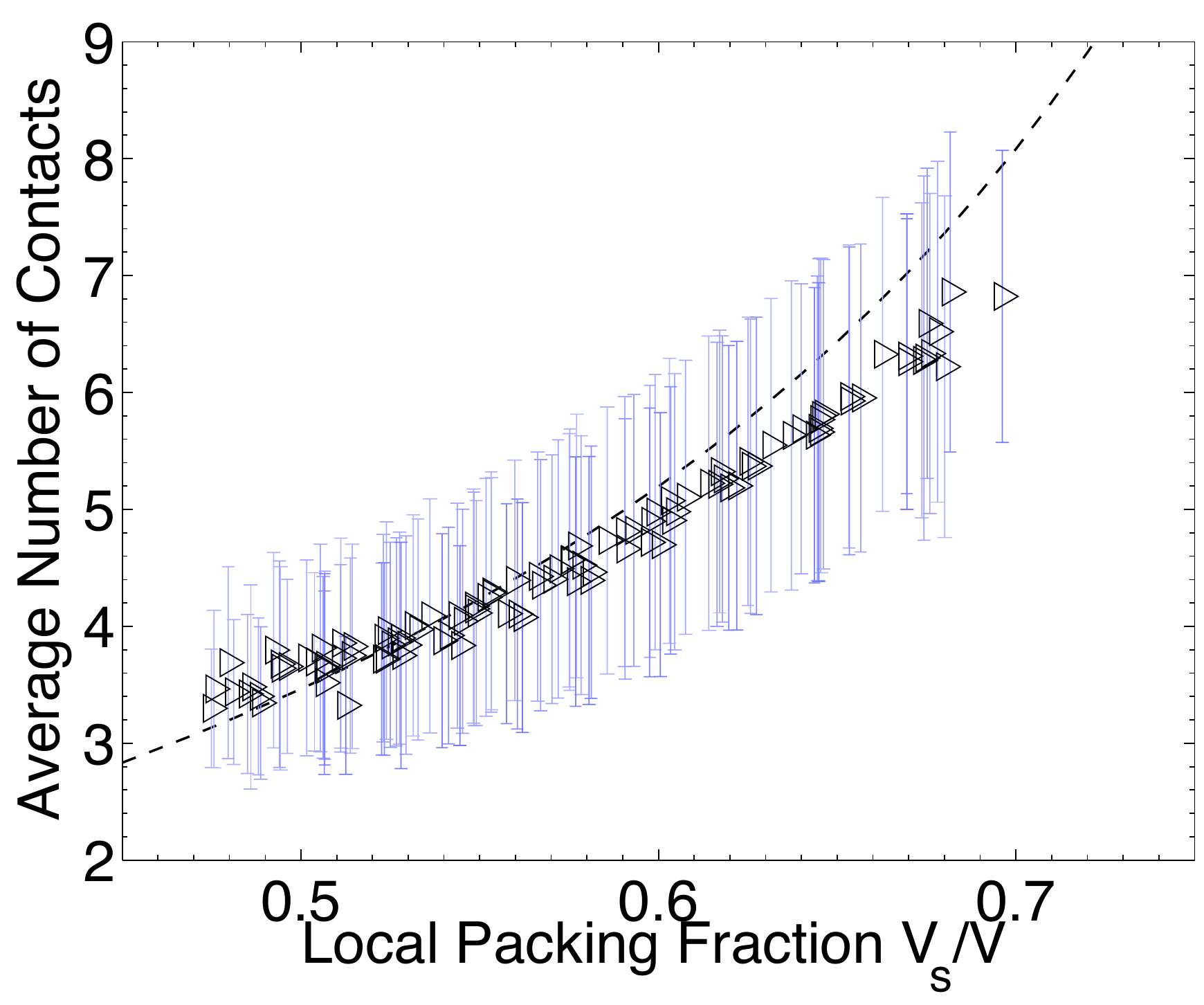} 
\includegraphics[width=.49\columnwidth]{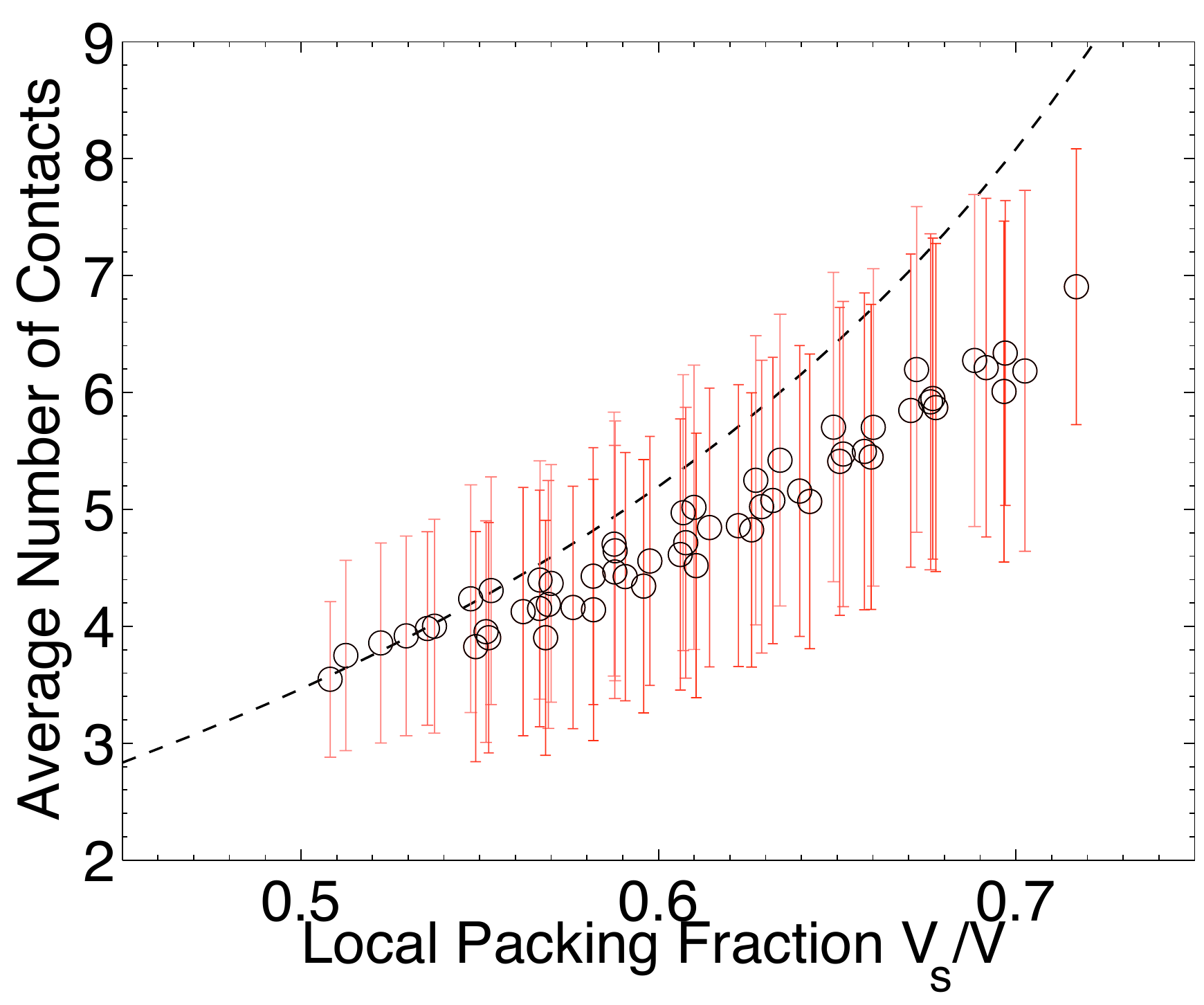}
\caption{ 
Local relation between the avenge number of neighbors in contact and the local packing fraction $\phi = V_s/V$.
Dashed line is the theoretical behavior from \cite{Song08}: $ \langle z \rangle= 2\sqrt{3}/(1/\phi-1) $.
Volumes are binned with into 24 bins between $V_{min}$ and $V_{Max}$.
The choice of binning has very marginal effects.
(top) Fluidized bed samples.
(bottom) Dry acrylic beads samples.
}
\label{f1}
\end{figure}

We have seen that the average number of contacts $z_c$ for the whole packing increases almost linearly with the packing fraction.
In a previous paper \cite{AsteLoacalGlobal06} it was pointed out that these two quantities are also locally related.
Fig.\ref{f1} reports the average number of contacts $\langle z \rangle$ for spheres at a given local packing fraction $\phi$.
The local packing fraction is calculated by computing the volume $V$ of the Voronoi cell around the sphere in the packing and computing the fraction of Voronoi volume occupied by the sphere, i.e. $\phi = V_s/V$ with $V_s = \pi d^3/6$ the volume of the sphere. 
Note that the global packing fraction ($\Phi$) is retrieved similarly form $\Phi = V_s/\rangle V \langle$.

An analytical approach in \cite{Song08}  proposes the following relation:
\begin{equation}
 \langle z \rangle= 2\sqrt{3}\frac{\phi}{1-\phi} \;\;.
 \end{equation}
 In Fig.\ref{f1} this analytical prediction is compared with the data for all the fluidized beads and dry samples.  
One can see that significant deviations are observed.

\section{Distribution of number of contacts}

\begin{figure}
\centering
\includegraphics[width=0.49\columnwidth]{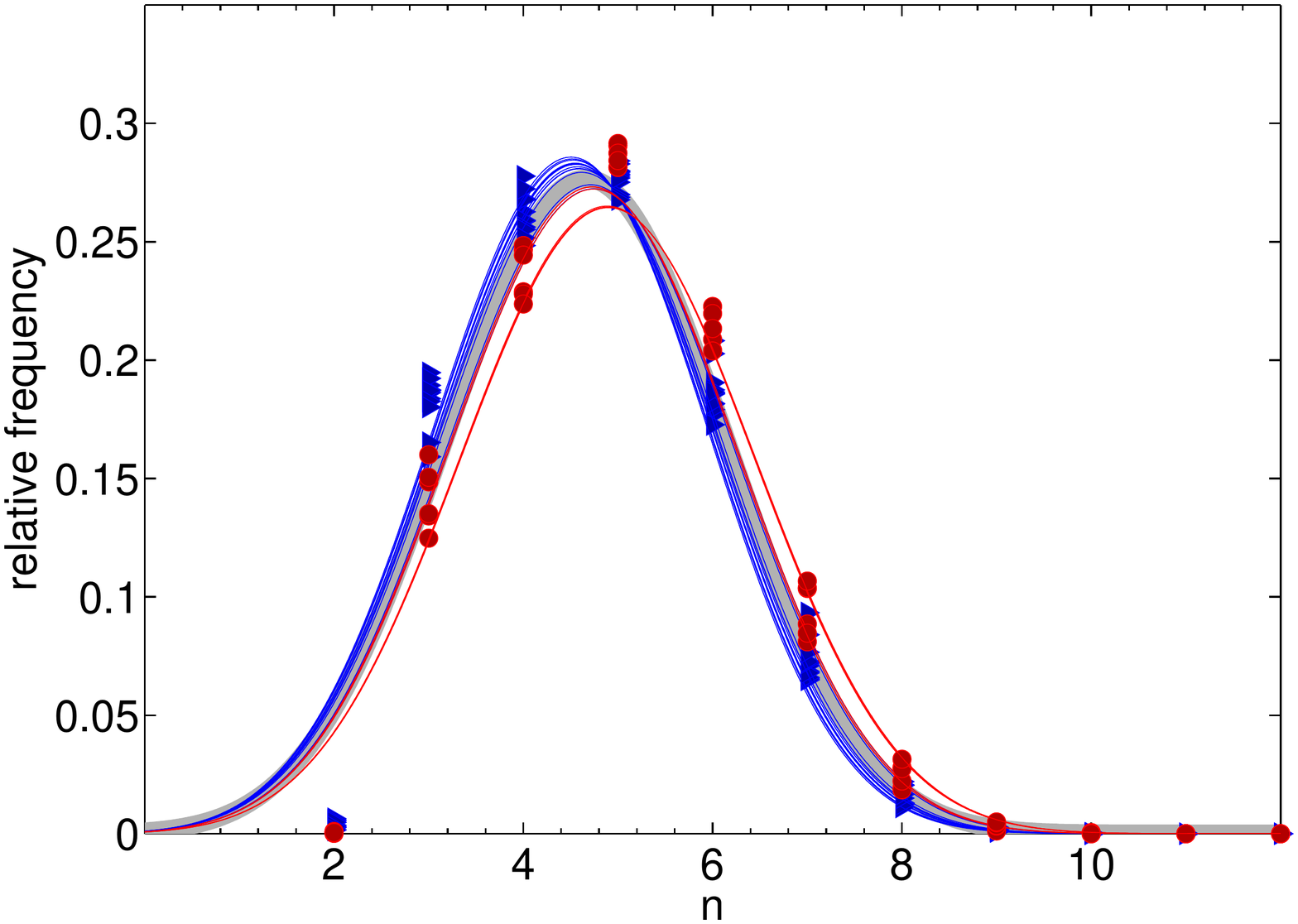}
\includegraphics[width=0.49\columnwidth]{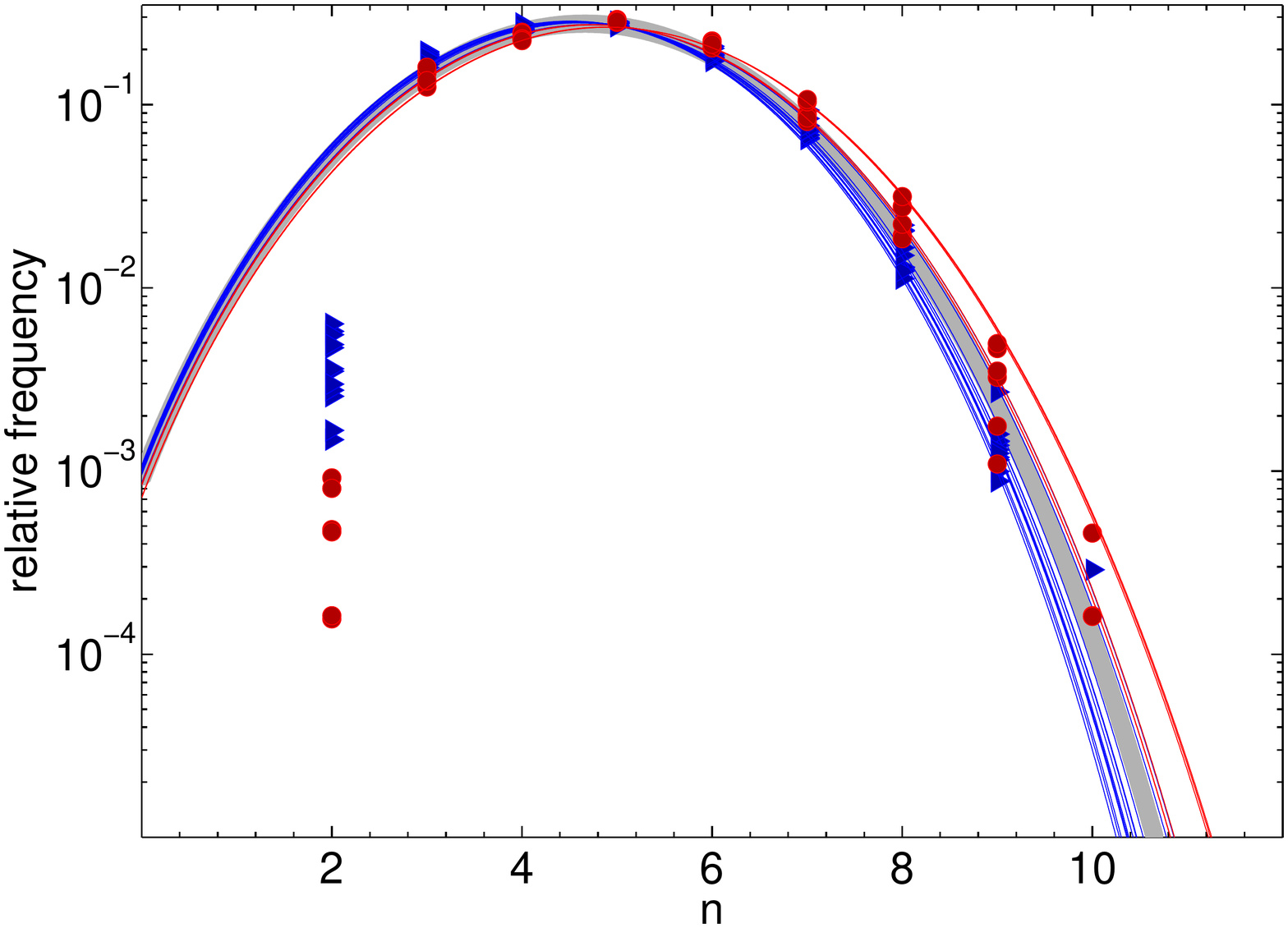}
\caption{  
Distribution of the neighbors in contact for all the experimental samples: fluidized bed ($\triangleright$) and dry acrylic beads ($\circ$).
The line is the function in Eq.\ref{z}.
}
\label{f.contactsDec}
\end{figure}

\begin{figure}
\centering
\includegraphics[width=0.49\columnwidth]{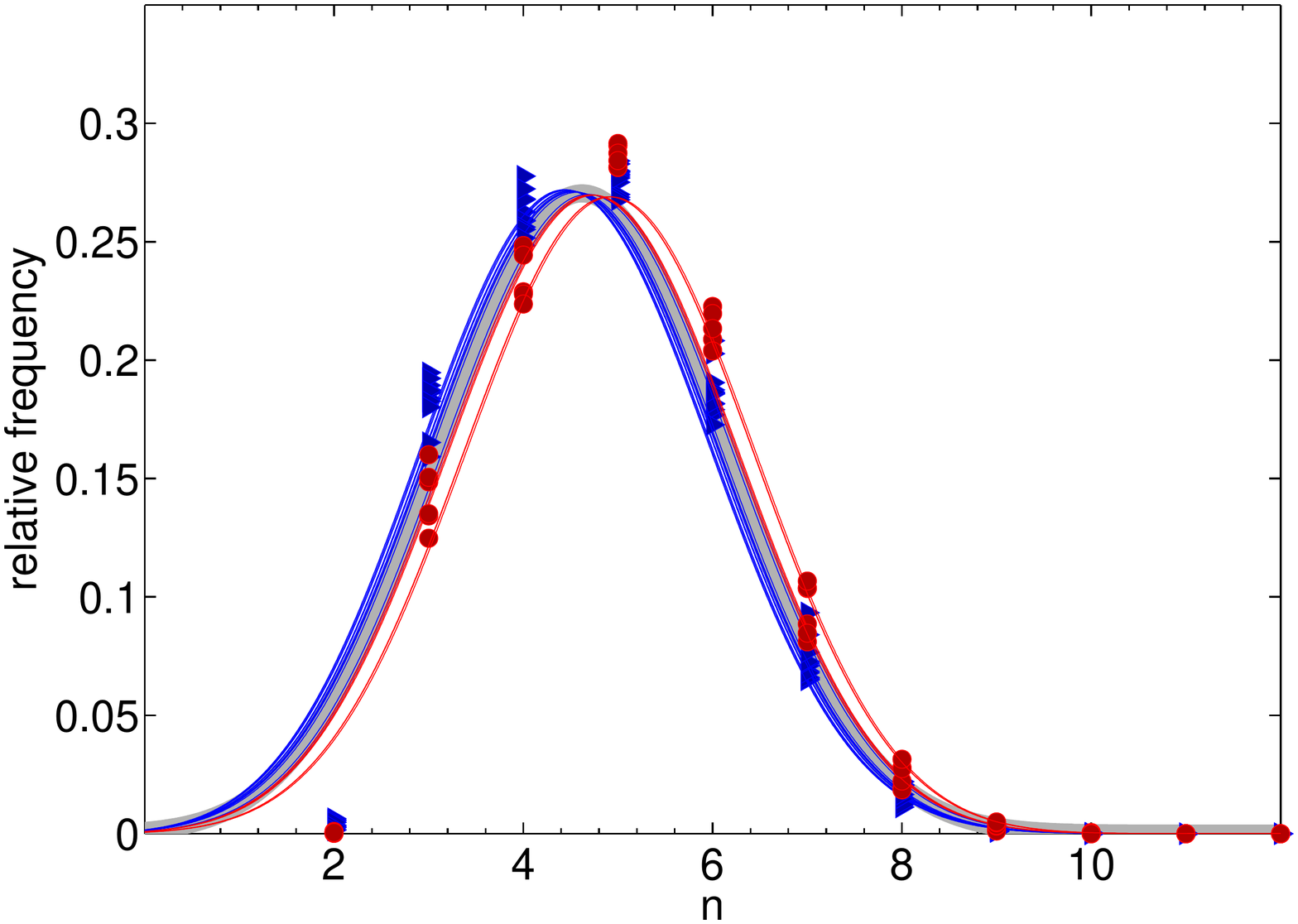}
\includegraphics[width=0.49\columnwidth]{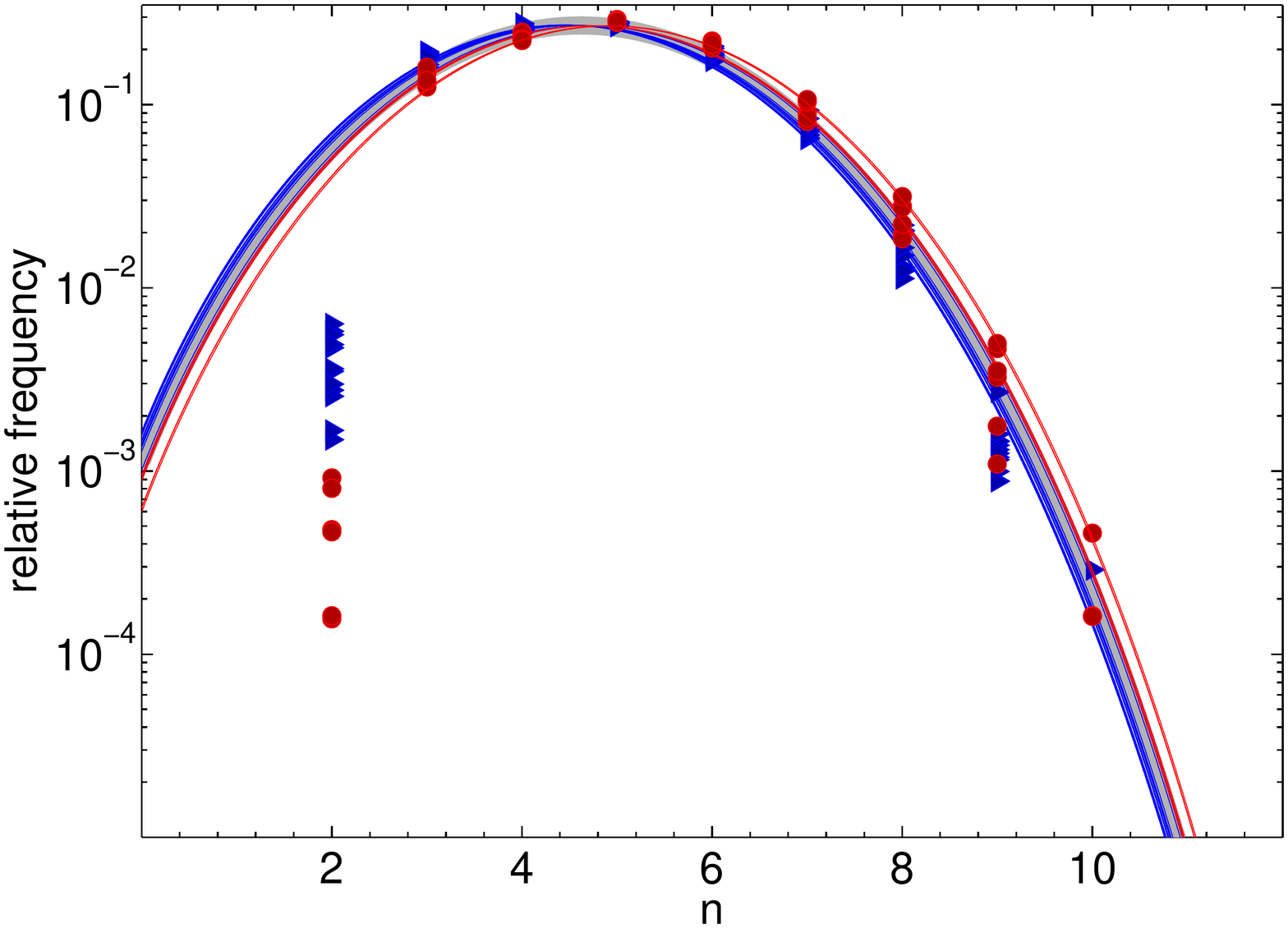}
\caption{  
Distribution of the neighbors in contact for all the experimental samples: fluidized bed ($\triangleright$) and dry acrylic beads ($\circ$).
The lines are the function in Eq.\ref{z1} with $z^* = 4/(2 - \sqrt{3}) \sim 14.9$ and the best fitting values of $\lambda$.
}
\label{f.contactsDec1}
\end{figure}

Figures \ref{f.contactsDec} and \ref{f.contactsDec1} report the distribution of the number of contact per sphere in the packings. 

Let us here introduce a  simple `free volume theory' applied to the spheres in contact with a given sphere can be used to estimate theoretically the probability distribution for the number of spheres in contact.
Let us consider the solid angle around a sphere $4\pi$ as a `volume'  which is locally filled by the neighboring spheres each occupying a solid angle $\Omega$.

In the `grancanonical' ensemble the partition function of a system of $z$ of such particles is $W(z)=(4\pi - \Omega)^z/z!$.
The probability to have a configuration with $z$ neighbors is therefore consistently $p(z)=W(z)/\sum_{z'} W(z')$ which can be written as:
\begin{equation}
p(z) = \frac{c}{\Gamma(z+1)} \left( 4 \pi \frac{z^* - z}{z^*} \right)^z   \;\;,
\label{z}
\end{equation}
where $z^* = 4\pi/\Omega$ is here considered a free fitting parameter and $c$ a renormalization constant such that $\sum_z p(z) = 1$.
In Fig.\ref{f.contactsDec1} the distribution of the number of neighbors in contact for all the samples are reported.
The lines are the plot of Eq.\ref{z} where the best fits for $z^*$ are used.
The resulting best fit values are in a narrow range between  14 and 15.
As one can see the theoretical prediction is very satisfactory describing well the overall behavior except for the probability for $z=2$ which is smaller than expected from Eq.\ref{z}.
This is not surprising,  indeed the theory ignores mechanical stability, but in reality grains with only two contacts are highly unstable and therefore less likely to be present.

The use of $z^*$ as a free fitting parameter is reasonable given that different packing strategies affect differently the likelihood  of filling the solid angle around a sphere.
However, in a strict mathematical sense, the free volume occupied by a touching sphere is  $\Omega = 2\pi (1-cos\pi/6)= \pi(2-\sqrt{3})$ and therefore $z^* = 4/(2 - \sqrt{3}) \sim 14.9$.
An alternative approach is considering $z^*$ fixed and instead taking into account of the different packing characteristics by using a sort of `chemical potential' that makes more or less likelihood for a sphere to be in touch with another sphere.
This alternative framework is equivalent to the `microcanonical' ensemble and the  `chemical potential' is a Lagrange multiplier $\lambda$ that controls the average number of neighbors in contact.
The expression for the probability becomes:
\begin{equation}
p(z) = c \frac{\left(z^* - z \right)^z}{\Gamma(z+1)}  \exp(-\lambda z) \;\;,
\label{z1}
\end{equation}
with $c$ a renormalization constant such that $\sum_z p(z) = 1$ and  $z^*=4/(2 - \sqrt{3})$. 
Fig.\ref{f.contactsDec1} shows that this alternative approach is also very effective in describing the distribution of the number of neighbors.
 

\section{Comparison with Isostatic Condition}

The samples considered here are in a mechanically stable state.
This applies to both the experimental samples that where measured at rest after pouring or fluid flow and to the DEM relaxed samples that where allowed a sufficiently long time to achieve the complete arrest of all grains. 
From a simple counting of the number of degrees of freedoms  with respect the number of constraints one can infer the minimum expected average number of contacts per grains that can allow mechanical stability.
Samples satisfying such a counting are generally referred as being at the isostatic condition.
One must stress that it has been mathematically proven that such a counting does not provide neither a sufficient nor a necessary condition for stability \cite{ppp}.
Nonetheless, the isostatic condition gives a good indication of the approximate value of contacts above which one can reasonably expect to find mechanically stable packings.
  
In our samples we have frictional spheres with rotational degrees of freedoms that interact with both normal and tangential forces.
The forces are the free variables of the problem that are subject to the constraints of zero total force and zero total torque acting on each grain.
Normal forces have directions determined by the position of the grains, therefore they count only for one scalar variable per contact, counting to: 
\begin{equation}
N_n = z_c \frac{N}{2}\;.
\end{equation}  
Tangential forces act in the plane between the two spheres, in general they have $D-1$ components (with $D=3$ the space dimensionality).
However, some contacts can be at the slipping threshold  where $F_t = \mu F_n$ and they cannot hold any extra tangential load in that direction, in this case the number of independent components is reduced by one becoming $D-2=1$.  
Overall the total number of variables associated with the tangential forces is: 
\begin{equation}
N_t= z_c \frac{N}{2} [(D-1)(1-q) + (D-2)q] = z_c \frac{N}{2} (D-1-q)\;,
\end{equation}
where $q$ is the fraction of contacts at the slipping threshold that cannot hold any extra tangential load, conversely $1-q$ is the fraction of contacts that are below the slipping threshold. 
The number of equations for force balance is: 
\begin{equation}
E_f = D N \;.
\end{equation}
Whereas the number of equations for torque balance is: 
\begin{equation}
E_t= \frac{D(D-1)}{2} N \;.
\end{equation}  
The balance between number of variables and number of equations is:
\begin{equation}
N_n + N_t = E_f + E_t
\end{equation}
resulting in the following condition for the average number of contact per grain at the `isostatic limit':
\begin{equation}\label{ISO}
	z_{ISO} = \frac{D(D+1)}{D-q}=   \frac{12}{3-q}  \;.
\end{equation}
It is clear that this implies $4 \le z_{ISO} \le 6$ depending on the fraction $q$ of contacts that are at the slipping threshold.
This number can be computed from the DEM relaxed samples and the values of $q$ for all the samples are reported in Tabs. \ref{t.1} and \ref{t.2}.
We observe that there is a sizable fraction of contacts at the slipping threshold with a larger fraction (about 15\%)  in the dry beads samples with respect to the fraction in the fluidized beds samples (about 3\%). 
This is due to the larger value of the friction coefficient in the fluidized bed samples. 
It is indeed clear that $q$ should depend on friction with the limit $q \rightarrow 1$ associated with the limiting case of infinitesimally small friction coefficients and conversely the limit $q \rightarrow 0$ associated with the limiting case of very large friction coefficients.
Overall we see from Tabs. \ref{t.1} and \ref{t.2} that all samples are `hyperstatic' with values of $z_c$ always considerably larger than the corresponding $z_{ISO}$.

\section{Critical cluster size}

In these packings mechanical stability is induced by the boundaries that literally hold all the sample in place.
Let us now imagine to extract an internal  cluster of spheres and investigate whether it can be mechanically stable under compression and/or shearing occurring at its boundaries. 
In this case, only the contacts between the grains inside the cluster contribute to stability.  
For such a cluster, the average number of internal-internal contacts per grains is smaller than $z_c$ because one must subtract the contacts with grains external to the cluster.
For larger clusters this number will converge towards $z_c$ but for small clusters it could be considerably lower than $z_c$ and even become smaller than $z_{ISO}$. 
The mechanically stability of  sub-parts of the packing can be inferred by dividing the entire sample in a grid of cubic cells of edge-size $l$ and looking at the average  internal-internal number of contacts inside each of these cells. 
The smallest cluster  size $l^*$ which correspond to  an average number of internal-internal  contact per grains larger or equal to $z_{ISO}$ can be considered as the `critical' size below which the system is locally in a mechanically unstable state and it is actually held into place by the presence of static neighbors.
Table \ref{t.2} reports the critical lengths $l^*$ for all samples. 
As one can see they vary in a range between $9$ to $12$ sphere diameters.




\begin{figure}
\centering
\includegraphics[width=0.49\columnwidth]{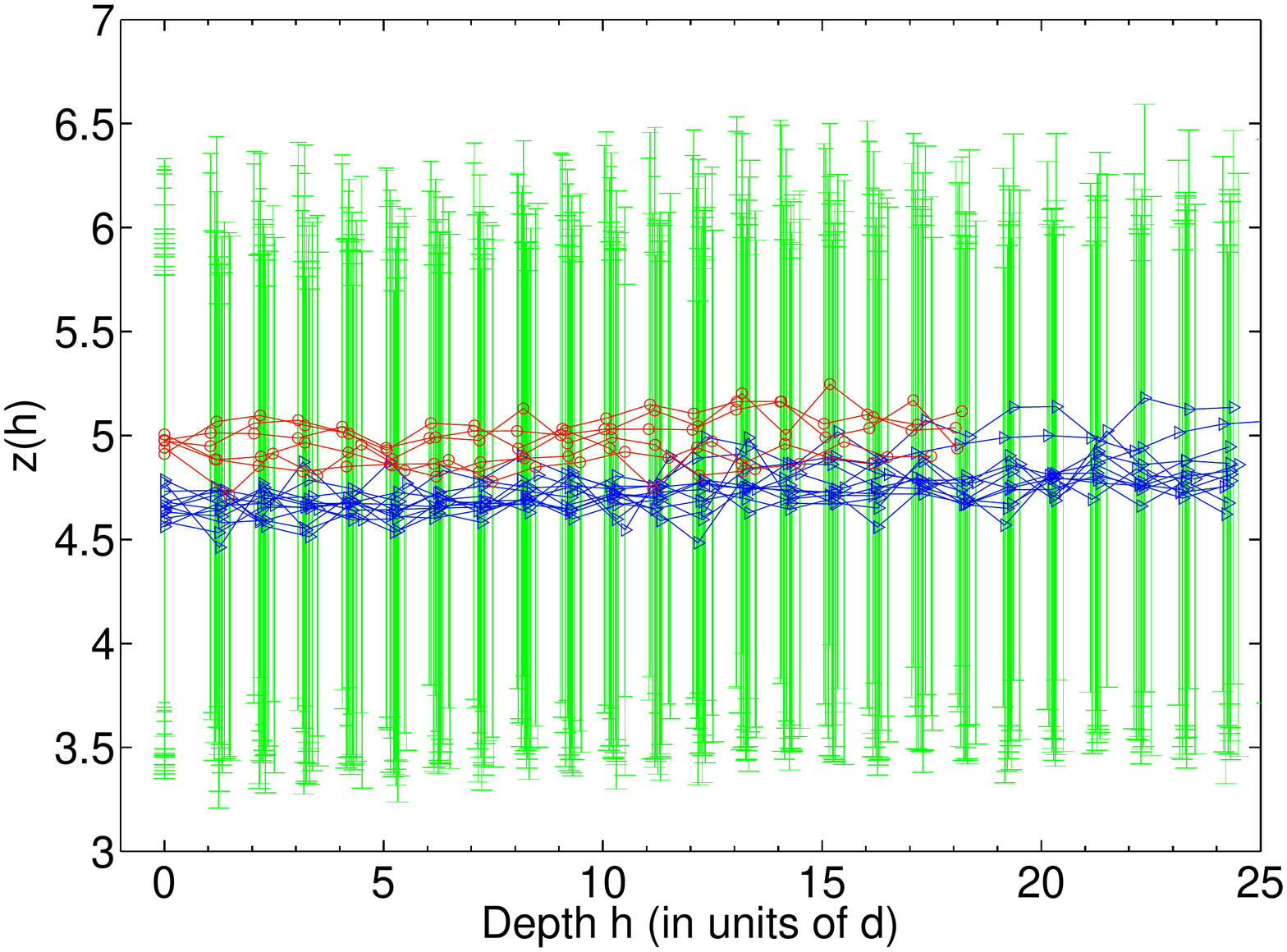} 
\includegraphics[width=0.49\columnwidth]{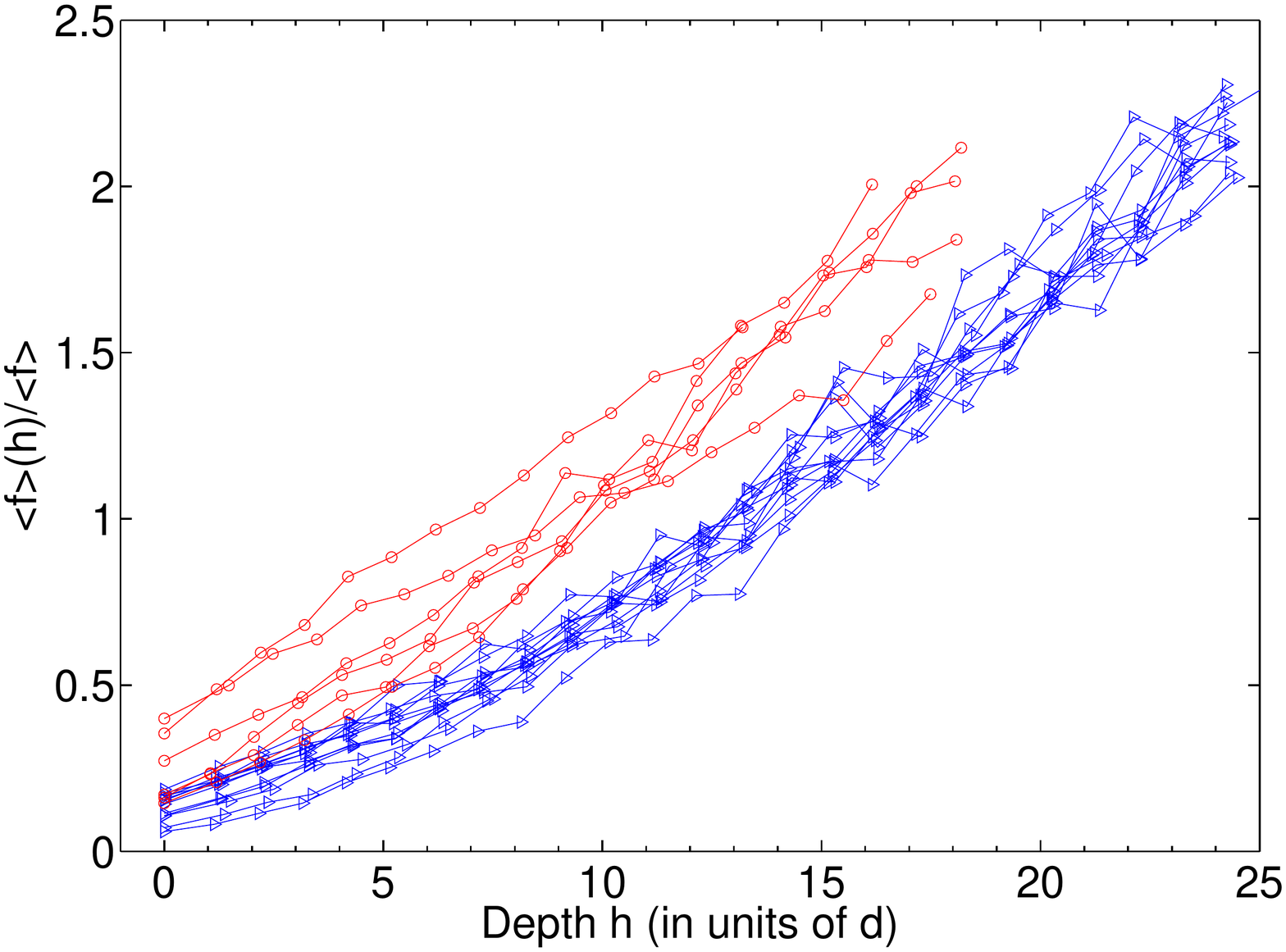}\\
\caption{
(left) Average number of contacts per grain vs. depth.
The error bars are represent the standard deviation.
(right) Average rescaled force vs. depth.  
}
\label{fvsh}
\end{figure}

\begin{figure}
\centering
 \begin{tabular}{cc}
\includegraphics[width=0.5\columnwidth]{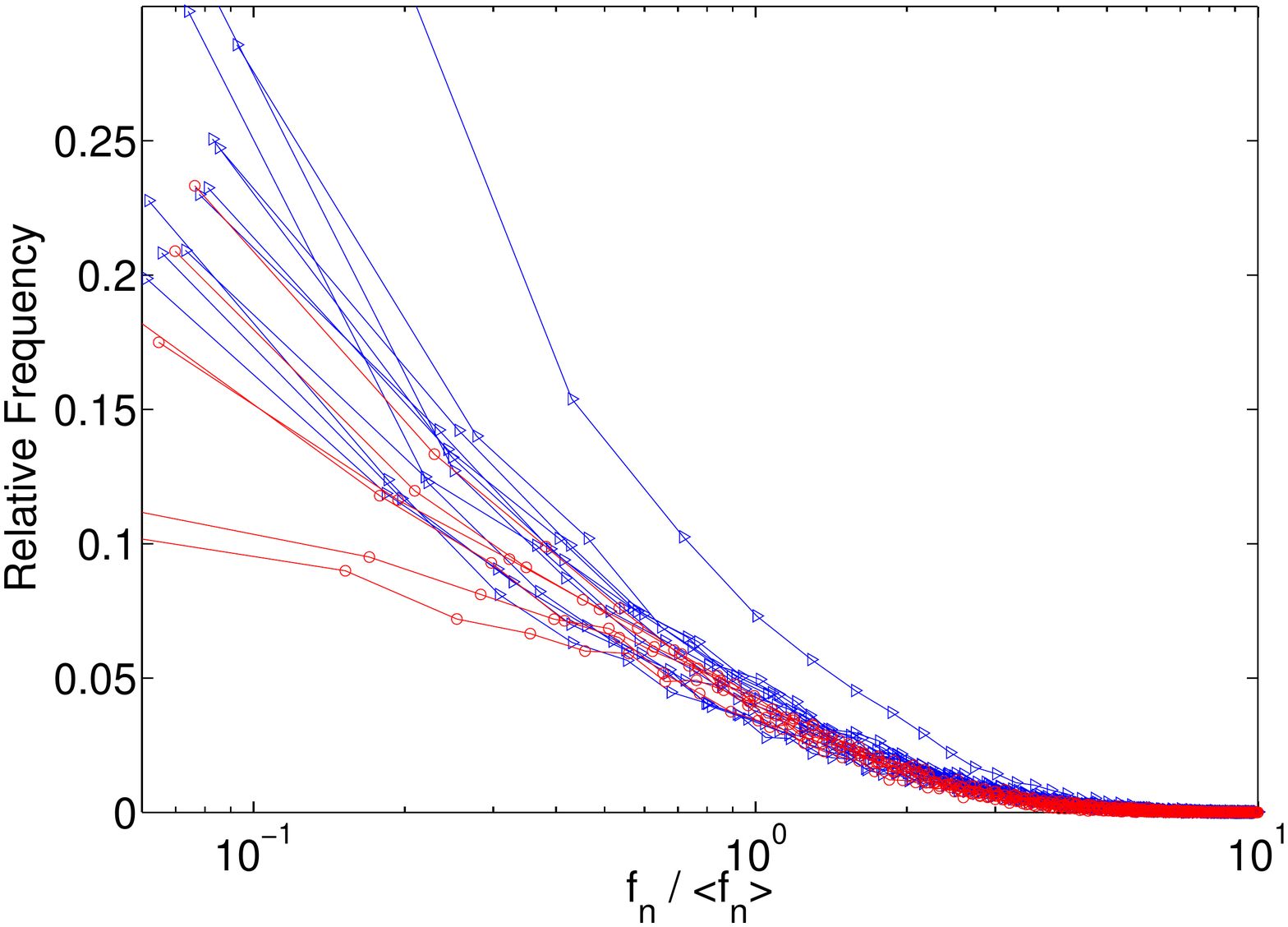}&\includegraphics[width=0.5\columnwidth]{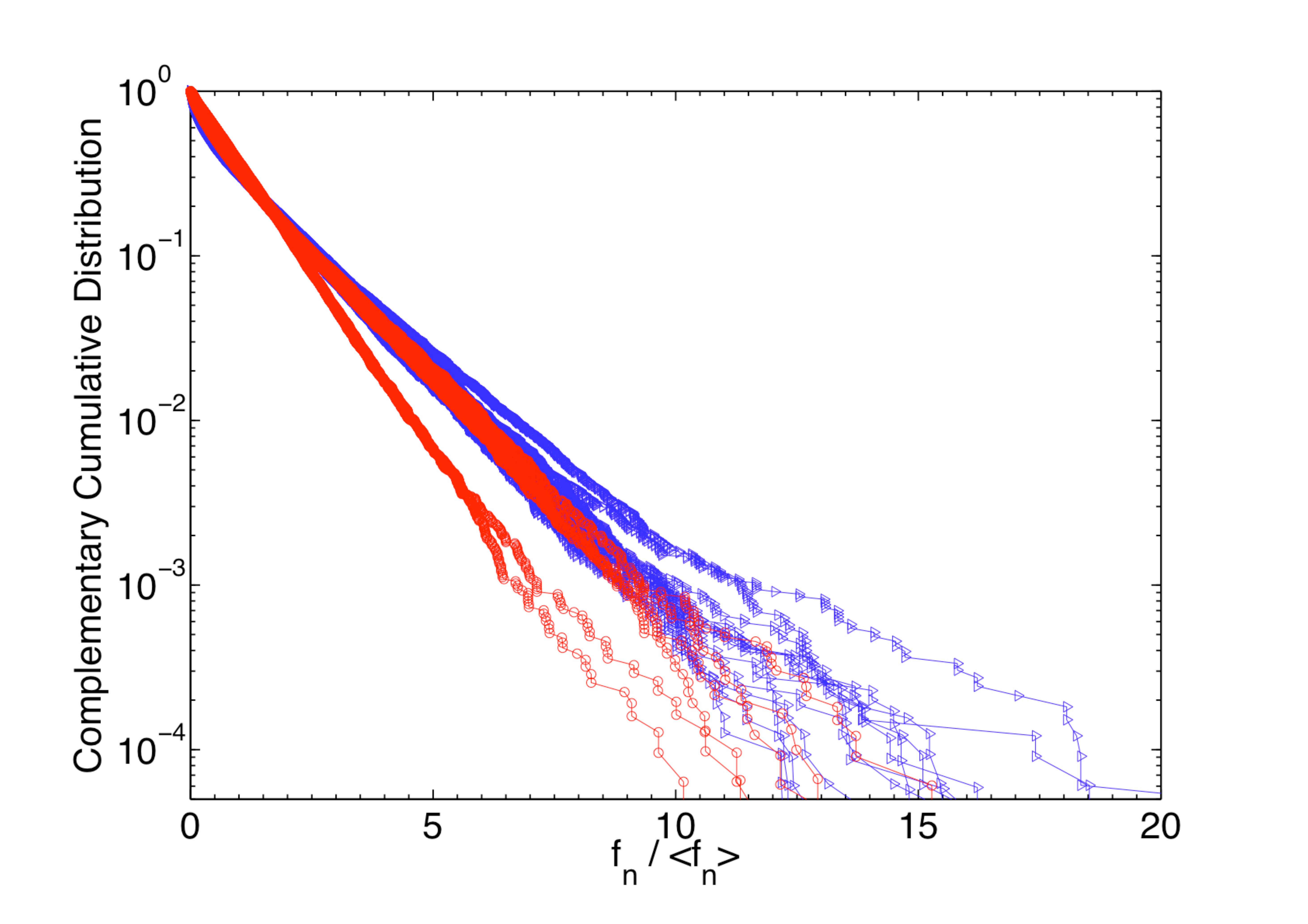}\\
\includegraphics[width=0.5\columnwidth]{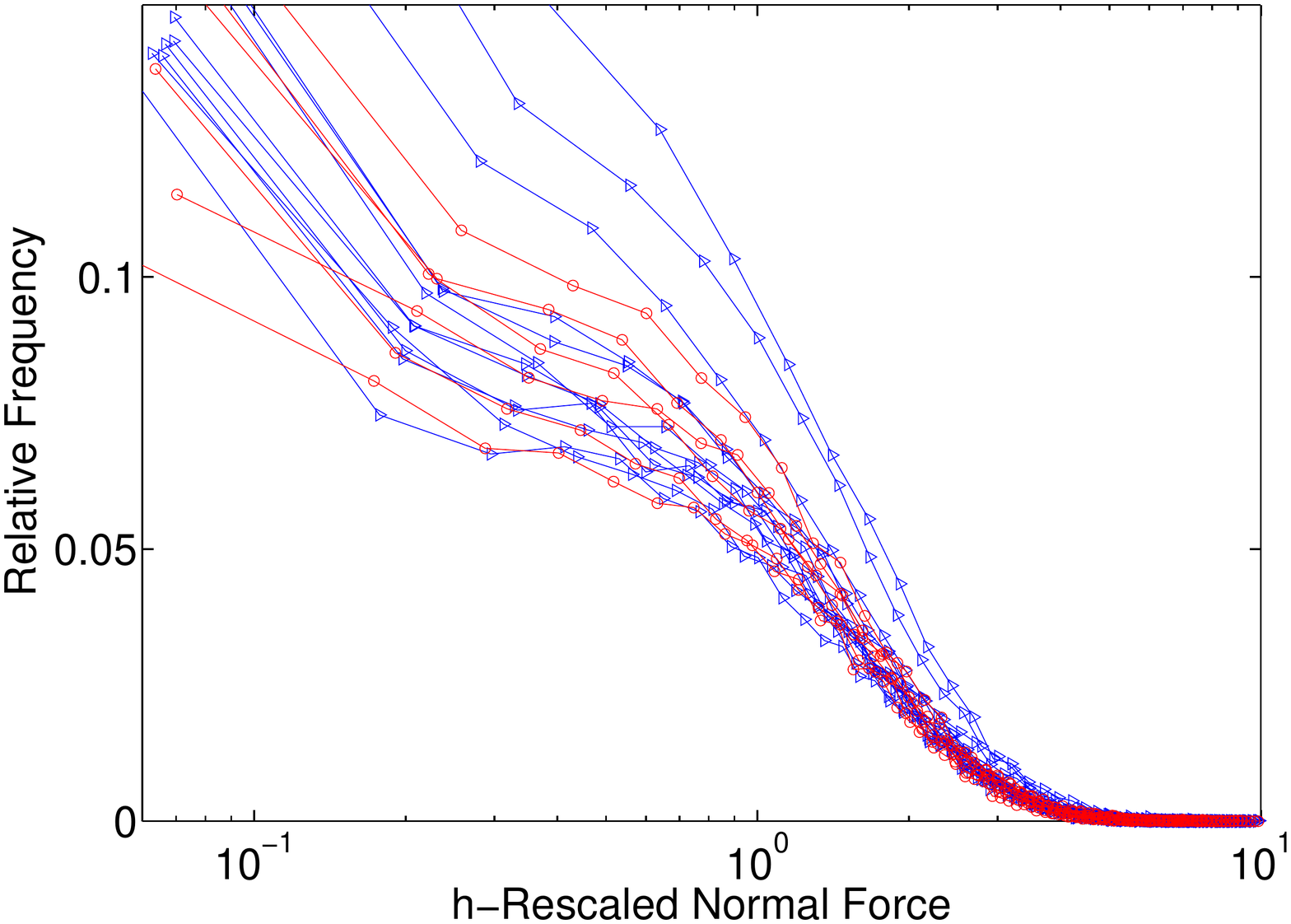}&\includegraphics[width=0.5\columnwidth]{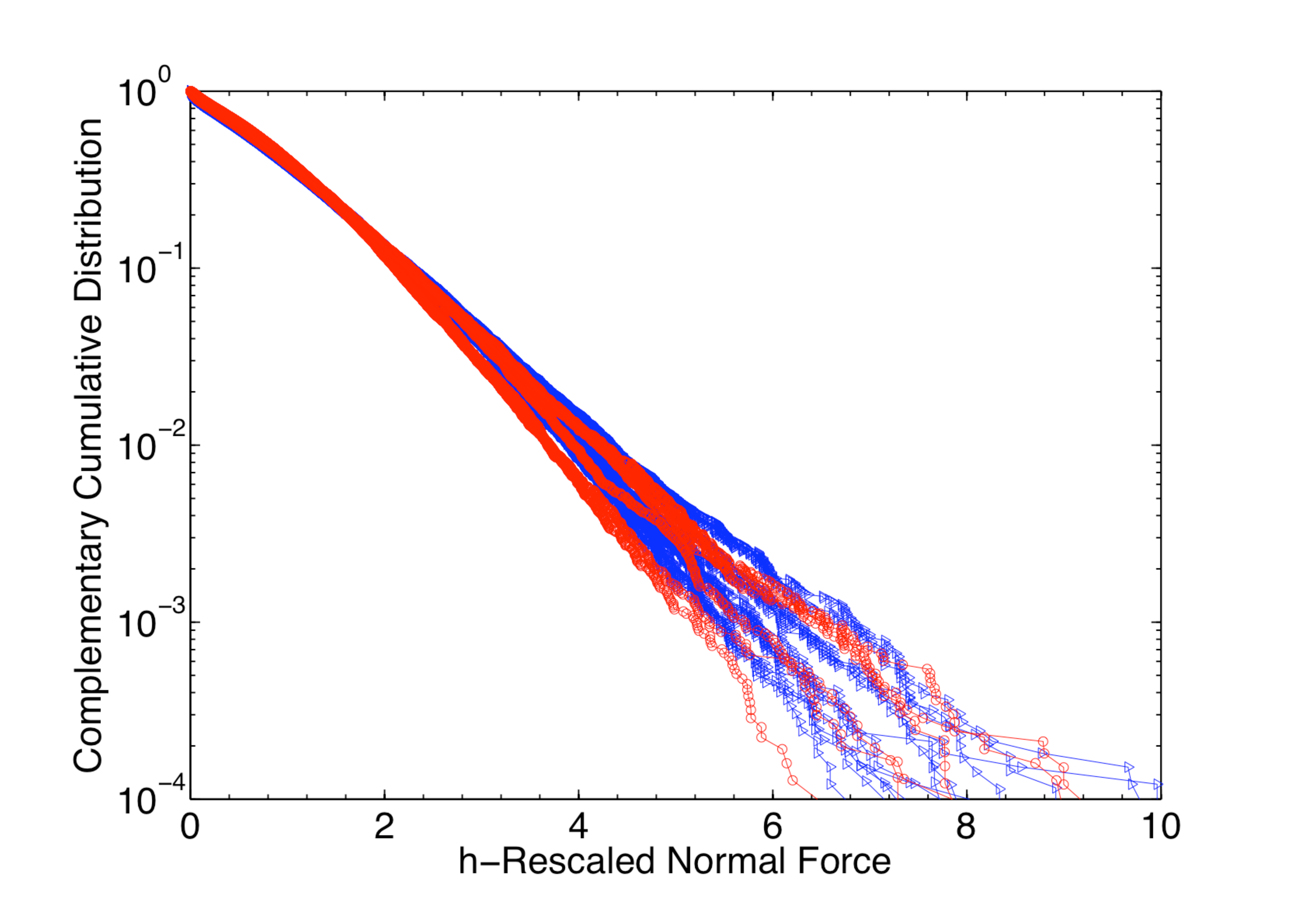}\\
\end{tabular}
\caption{  
Distribution of normal forces  inside the sample.
(above) Distribution of the forces rescaled by the average force in the sample.
(below) Distribution of the forces rescaled by the average force in each layer at different depths.
}
\label{fdistr}
\end{figure}

\section{Distribution of forces}

Understanding how stress distributes in these systems is of great relevance to unveil the mechanisms underneath mechanical stability and the jamming transition in these complex materials.

Let us first note that the topological properties of these samples are rather insensitive to the relative depth of the grains.
An example is given in Fig.\ref{fvsh} where the average number of neighbors and the standard deviation of the distribution are plotted for several layers at different depths, showing that these quantities rest almost unchanged.  
A different behavior is instead observed for the average normal force in each layer.
Increasing the depth the force increases due to the larger weight of the grain above. 
We will take into account this effect in the following discussion.

Figure \ref{fdistr} reports the distribution of the normal forces between grains measured in an internal part of the sample 4 sphere diameters away from the boundaries.
The log-linear scale reveals a linear behavior in the tail of the distribution indicating an exponential decay. 
We observe that the rescaling of the force with the mean force measured in a layer at the corresponding height (the one reported in Fig.\ref{fvsh}) results into a cleaner collapse of all sample distributions.

\begin{figure}
\centering
 \begin{tabular}{cc}
\includegraphics[width=0.5\columnwidth]{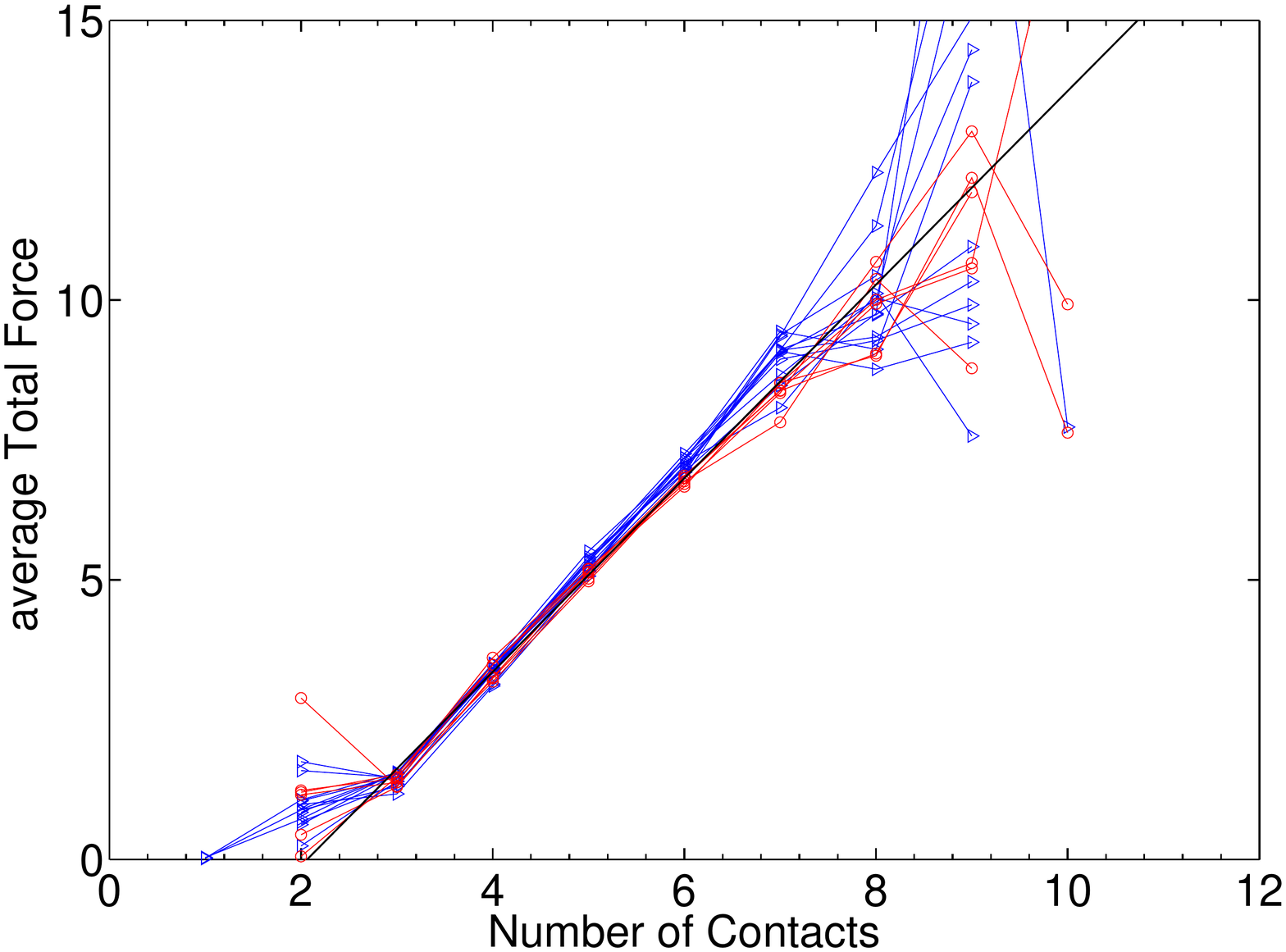}
\includegraphics[width=0.5\columnwidth]{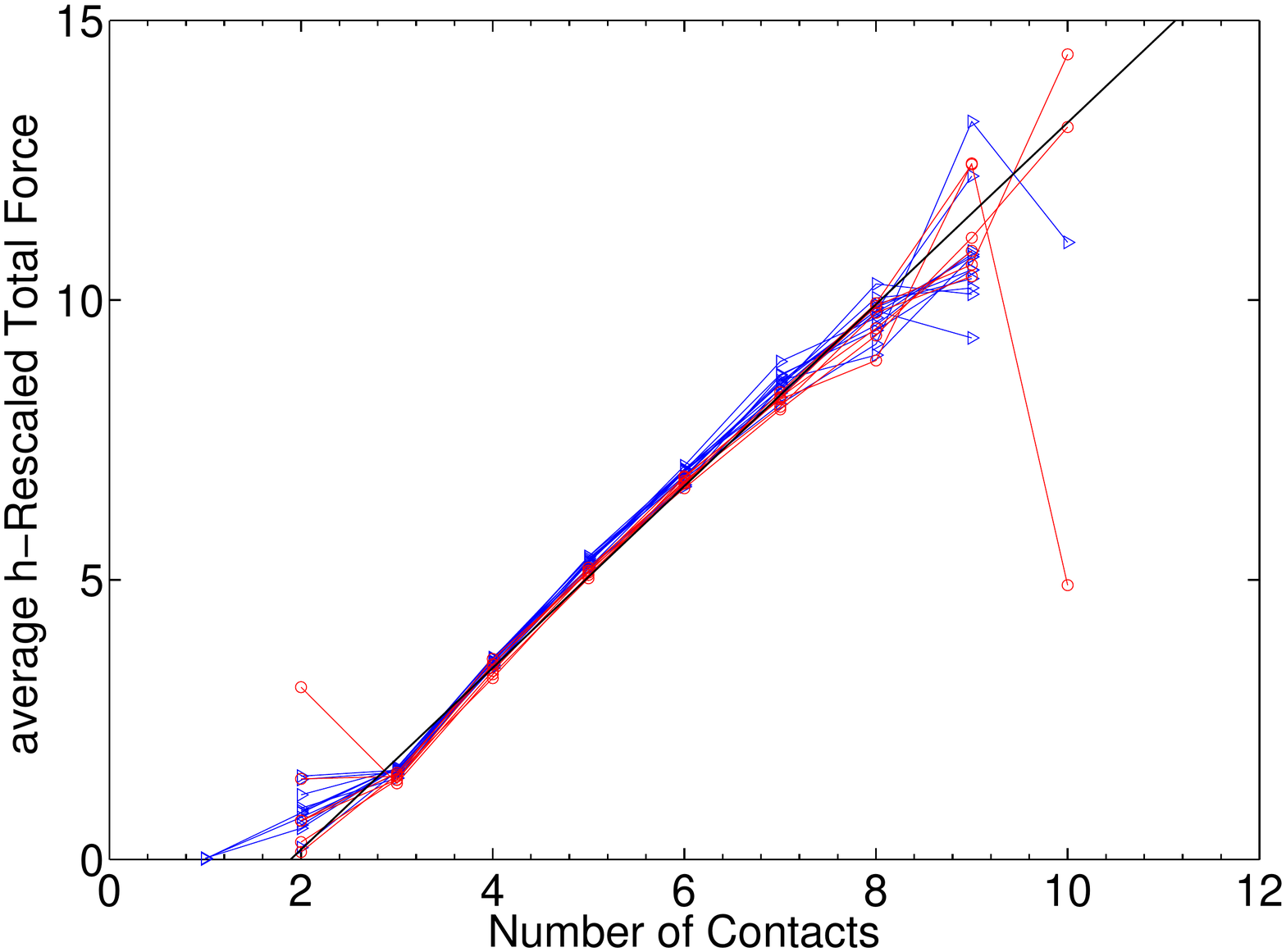}
\end{tabular}
\caption{  
Average total force vs. number of contacts per sphere. }
\label{fContactForce}
\end{figure}

In Fig.\ref{fContactForce} we plot the average total normal force acting on a sphere with a given number of contacts $z_c$.
We observe a very neat liner increase with $z_c$ which might indicate absence of correlations. 

\section{Local pressure}

\begin{figure}
\centering
\includegraphics[width=0.49\columnwidth]{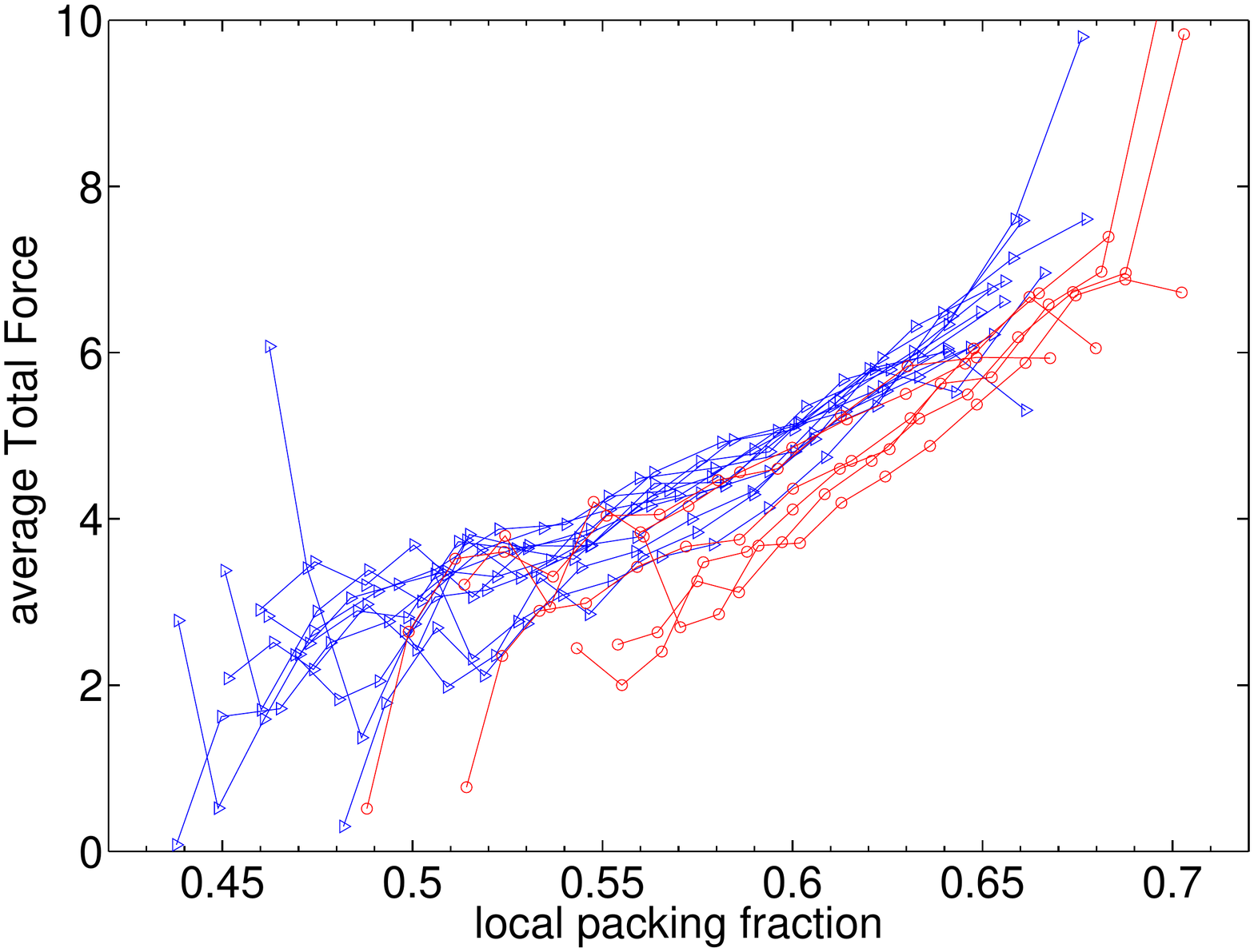}
\includegraphics[width=0.49\columnwidth]{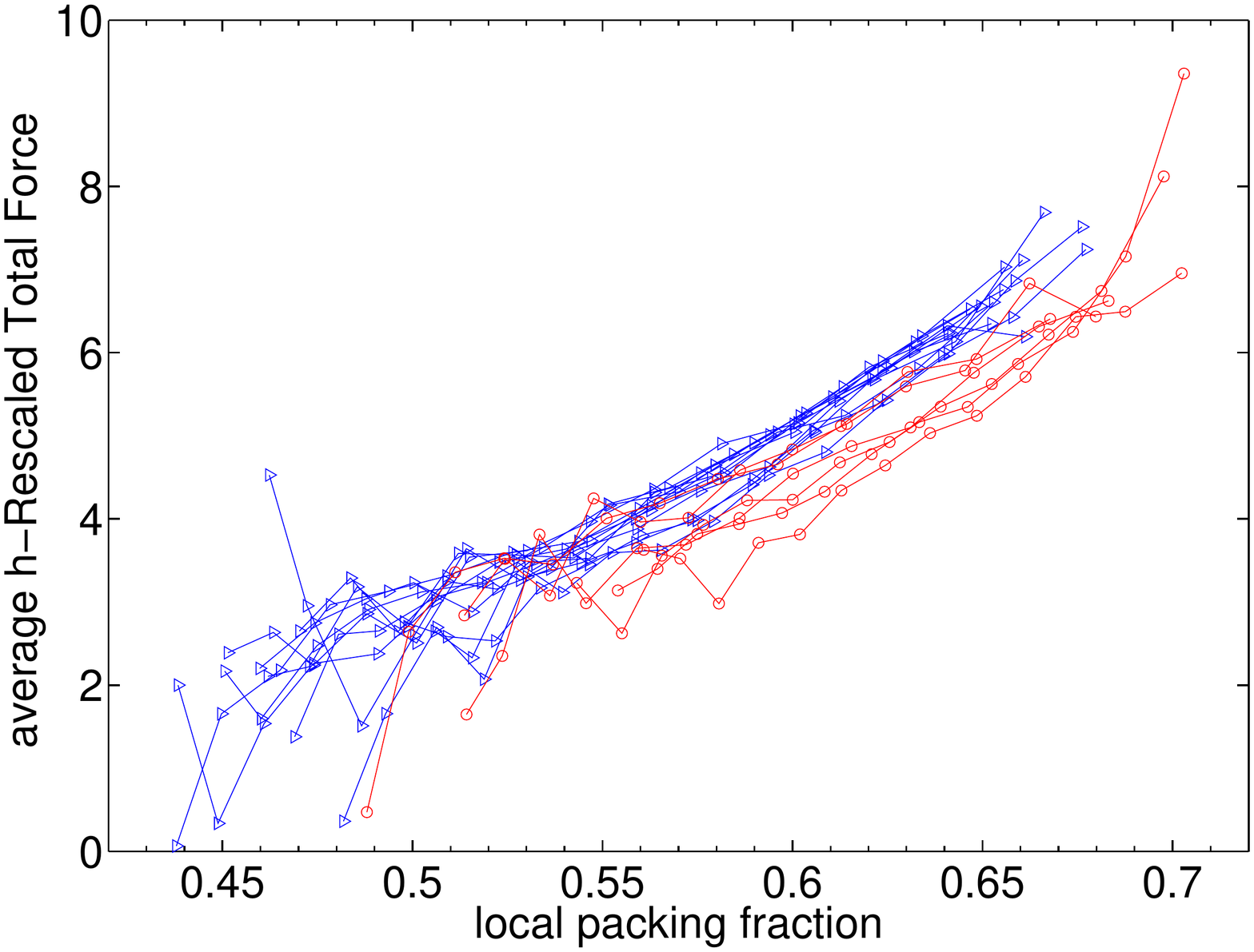}
\caption{  
Total rescaled force acting on a spheres with different local packing fractions.
(left) The total force is rescaled to the sample average.
(right) The total force is rescaled to the layer average.
}
\label{fForce}
\end{figure}

Investigating the relation between local packing fraction and force acting on a grain is a very interesting issue that might give insights on mechanical stability.
In Figure\ref{fForce} we report the total normal force acting on a given sphere as function of the local packing fraction i.e. $\phi = V_s/V$.
It is clear that there is a monotonic relation with densely packed grains carrying larger forces.
The figure reports two rescalings with respect to the average force in the whole sample and in each layer.

\begin{figure}
\centering
 \begin{tabular}{cc}
\includegraphics[width=0.5\columnwidth]{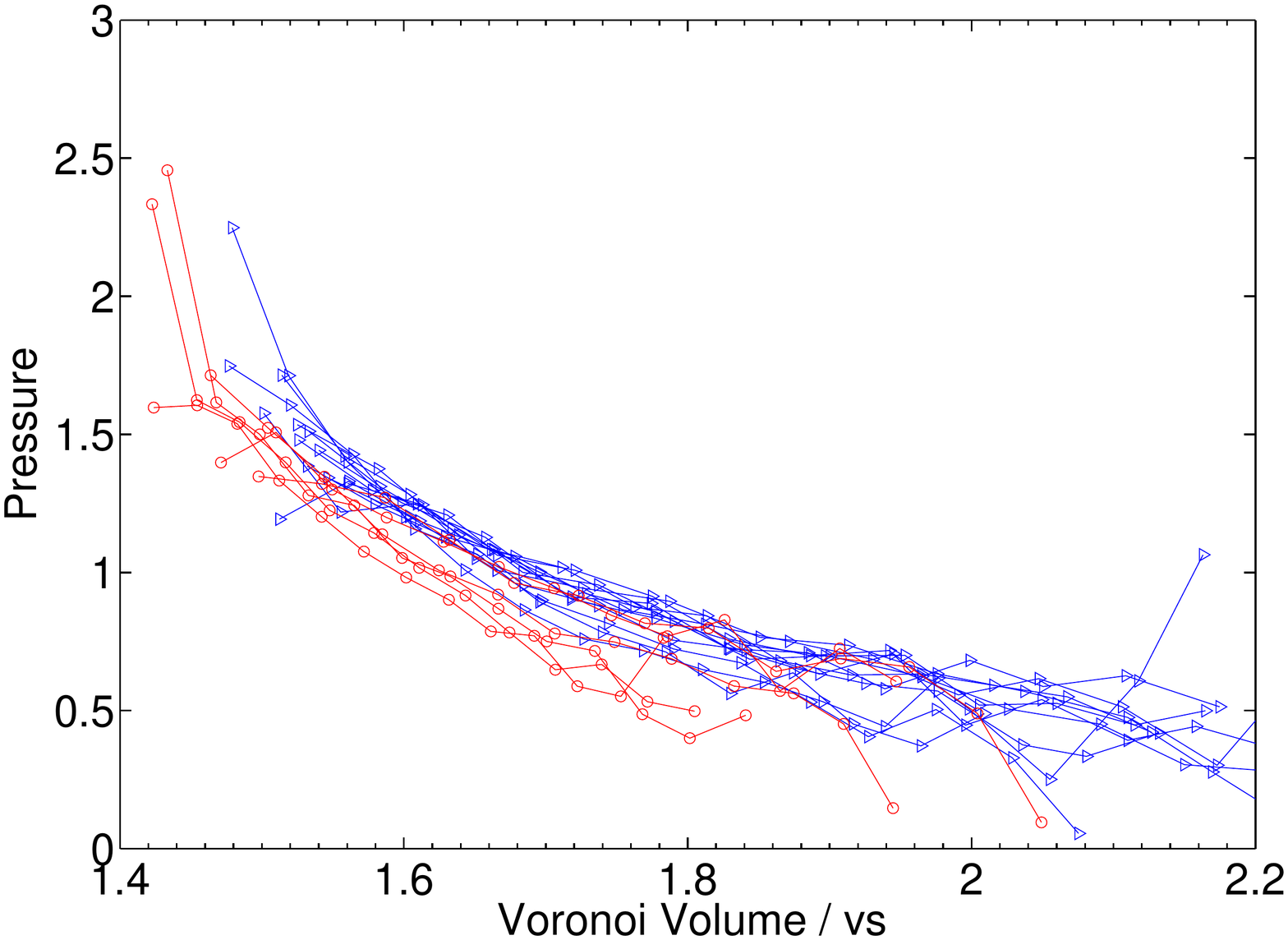}
\includegraphics[width=0.5\columnwidth]{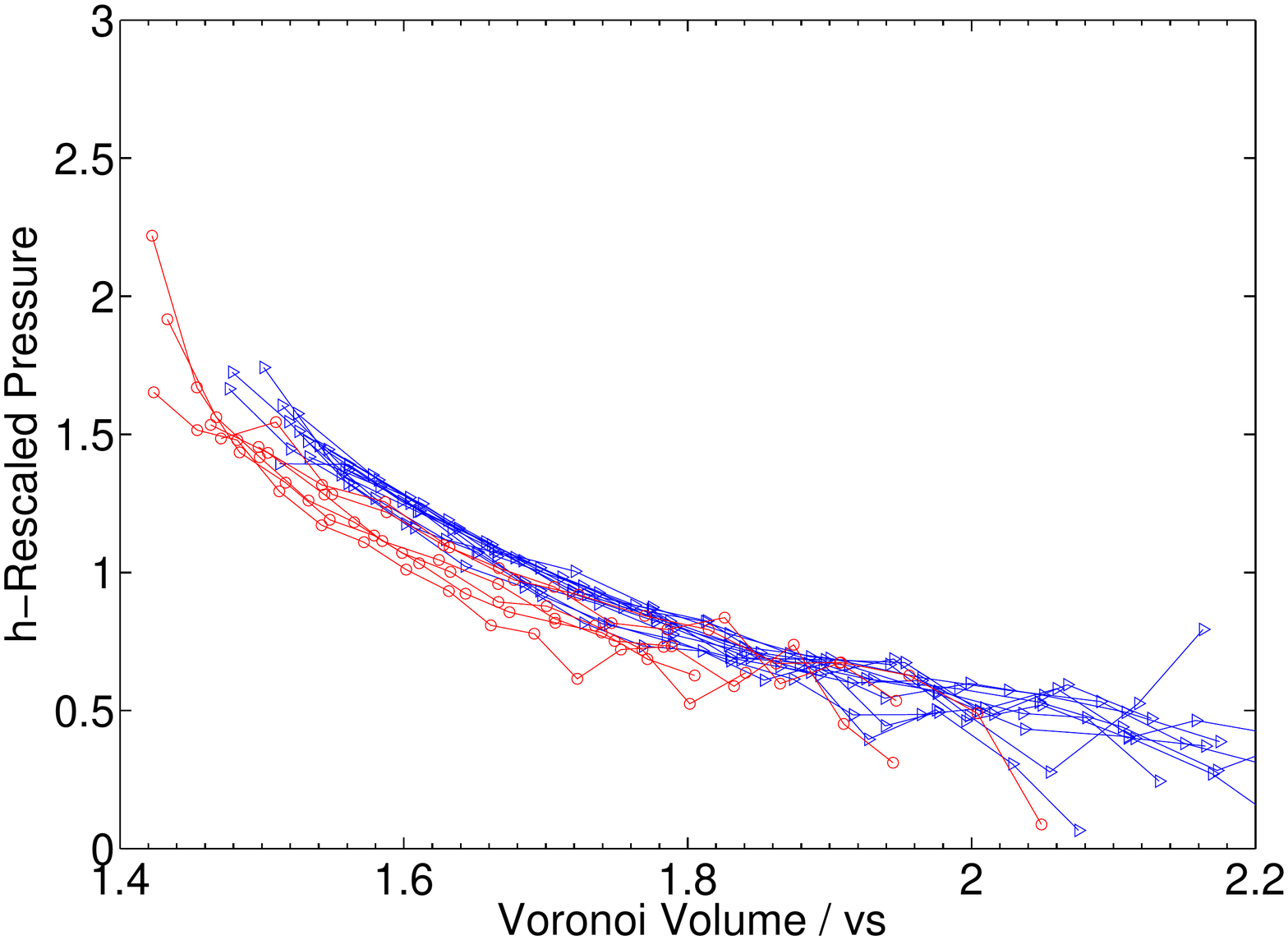}
 \end{tabular}
\caption{  
Average pressure (total rescaled force / Voronoi area) acting on a spheres with different Voronoi volumes.
(left) The total force is rescaled to the sample average.
(right) The total force is rescaled to the layer average.
}
\label{fPressure}
\end{figure}

A similar measure is reported in Fig.\ref{fPressure} where instead of the force the `pressure' is considered.
In this context the pressure is defined as the total normal force acting on a sphere divided by the total area of the corresponding Voronoi cell. 
Data are renormalized by dividing by the average force (both in the whole sample and in the layers) and by multiplying by $d^2$ (to renormalize to the case of unit diameter spheres).
The result is analogous to what is observed for the forces, with the smaller Voronoi volumes (larger $\phi$) carrying larger pressures. 
 

\section{Conclusions and Outlook}

In this work we presented results obtained from a newly developed  `virtual laboratory' platform which combines direct experimental imagining data from XCT with DEM simulation tools.
We have shown that such a combination is very powerful providing virtual samples which are geometrically identical to the experimental ones but allow us to access to physical information that would not be obtainable from experiments alone. 
For instance, we have been able to greatly improve our estimate of the number of grains in contact and to study accurately the region of near contacts.
In particular, we pointed out that the exponential law for the growth in the number of near neighbors with the radial distance is well followed across all samples but the exponent is not universal varying from $0.7$ at low packing fractions to $0.5$ for the densest sample at $\Phi \simeq 0.64$.
We have discussed how close our packings are to isostaticity and how a critical size of around ten grain diameters emerges as minimal mechanically-stable cluster.
The relation between average number of contacts and packing fraction has been investigated both globally and locally and it has been compared with analytical predictions.
The distribution of the number of contact per grain has been investigated and we introduced a simple meaningful model based on free `particles' sharing the solid angle around a sphere which turns out to reproduce remarkably well the observed distributions.
We have been able to access the complete set of forces on each individual grain and we have shown that the normal forces well follow an exponential distribution.
We measured the local pressure on each Voronoi cell and we uncover a liner relation  between the sum of the normal forces and the number of contacts on each bead. 
In future work, we will further examine the dynamic properties of packings, considering the system under external loadings and under shear, combining high speed tomography and other bulk imaging techniques with DEM during dense granular flow and at the jamming transition. The role of grain shape in granular systems is also of great interest \cite{Donev04c,Delaney2008} and an extension of this technique to consider three-dimensional non-spherical grains will also be pursued.

\section{Acknowledgements}
Many thanks to T.J. Senden and M. Saadatfar for the tomographic data. 
This work was partially supported by the ARC discovery Project No. DP0450292 and Australian Partnership for Advanced Computing National Facilities (APAC).


\end{document}